# Photon wave functions, wave-packet quantization of light, and coherence theory

**Brian J Smith**[1,2] **and M G Raymer**[2]

[1]Clarendon Laboratory, Oxford University, Parks Road, Oxford, OX1 3PU, UK
[2]Oregon Center for Optics and Department of Physics, University of Oregon, Eugene, Oregon, 97403, USA

Email: `b.smith1@physics.ox.ac.uk`

**Abstract.** The monochromatic Dirac and polychromatic Titulaer-Glauber quantized field theories (QFTs) of electromagnetism are derived from a photon-energy wave function in much the same way that one derives QFT for electrons, that is, by quantization of a single-particle wave function. The photon wave function and its equation of motion are established from the Einstein energy-momentum-mass relation, assuming a local energy density. This yields a theory of photon wave mechanics (PWM). The proper Lorentz-invariant single-photon scalar product is found to be non-local in coordinate space, and is shown to correspond to orthogonalization of the Titulaer-Glauber wave-packet modes. The wave functions of PWM and mode functions of QFT are shown to be equivalent, evolving via identical equations of motion, and completely describe photonic states. We generalize PWM to two or more photons, and show how to switch between the PWM and QFT viewpoints. The second-order coherence tensors of classical coherence theory and the two-photon wave functions are shown to propagate equivalently. We give examples of beam-like states, which can be used as photon wave functions in PWM, or modes in QFT. We propose a practical mode converter based on spectral filtering to convert between wave packets and their corresponding biorthogonal dual wave packets.

**1. Introduction—Photon Wave Mechanics**

There are still many puzzling aspects of the nature of light. A central point to understand is the distinction between a corpuscular viewpoint and a field viewpoint of light. In the old days, this was called wave-particle duality, but this phrase does little justice to the subtle issues involved. A key question is, can we view light as being comprised of particles called *photons*, or must one view light as a field, and the "number of photons" only as the name we give to quantum states of the electromagnetic field [1]? Certainly, one can create single-photon wave packets, which are more or less localized in space-time, and we can describe them using standard quantum field theory (QFT) [2-4]. In some papers, authors ask questions such as, which path did the photon take through my interferometer? What slit did the photon go through? With sophistication, such questions can properly refer to alternative quantum amplitudes that contribute to the final amplitude for detection. These questions, however, presuppose that we know how to write down formulas for photon wave functions to represent these alternatives, along with a proper quantum wave equation for the photon wave function. Most papers fail to do that, yet many continue to use the photon-as-particle language loosely or sometimes even sloppily. In cases where we care only about which beam a photon follows as it traces its way through an interferometer made of beam splitters and mirrors, such a coarse language is probably fine [5, 6]. On the other hand, when a problem involves light diffraction, ultrashort pulses, or other spatially complex phenomena, one needs to use a more refined theory based on a photon wave equation.

It is known that one can describe single-photon states of light using a photon-as-particle viewpoint, specifying the photon wave function (PWF). We call this approach *photon wave mechanics* (PWM). Nevertheless, not all quantum optics researchers are well versed in the techniques for handling single-photon wave packets and photon wave functions. Also, some of the deeper connections between photon wave functions, quantized-field wave packets, and optical coherence theory have not been previously reported. These concerns motivate the present paper.

A main theme of this article is that there is utility in being able to switch correctly between a photon viewpoint and a field viewpoint. We begin by reviewing and extending the QFT of photon wave packets introduced by Titulaer and Glauber (T-G) [7]. We then briefly review the theory of the "energy-density photon wave function" in coordinate space, which has developed over the past dozen years [8-11], and we extend it in several ways. We show that both the quantized field theory of light developed by Dirac [12], based on monochromatic modes, as well as its generalization to non-monochromatic modes developed by Titulaer and Glauber, can be derived directly from the photon-as-particle viewpoint. This actually provides a "derivation" of the Maxwell equations, starting from fundamental principles. One does this by considering the relativistic particle kinematics of a single photon, and finding a formulation for a single photon that is analogous to the Dirac equation for an electron, which turns out to have the same form as Maxwell's equations. We then quantize single-photon theory to create a quantum field theory of light. The derivation parallels that of Dirac for the electron and its quantum field [13, 14]. A key difference between the electron and photon derivations has to do with the famous localization problem for the photon [15]. Whereas (non-relativistic) electrons can be in a position eigenstate, at least in principle, a photon cannot. On the other hand, the energy density of the electromagnetic field in free space can be expressed as a local quantity, $E^2(x)+c^2B^2(x)$. Therefore, as well argued by Bialynicki-Birula [8-10] and by Sipe [11], for photons it is best to adopt a wave function whose modulus squared is the photon's mean energy density, rather than being a position probability density, as is the case for electrons. We call this the *mean-energy-density wave function* or the Bialynicki-Birula-Sipe (BB-S) wave function and its equation of motion the *photon wave equation*. A connection of PWM to experiments can be seen in the authors' determination of transverse spatial PWFs at the single-photon level [16].

The wave functions (states) of a single photon, when treated as a particle-like object, are found to be equivalent to the mode functions of the quantized electromagnetic field. This field is conveniently expressed in terms of the complex electromagnetic field, $\mathbf{E}+ic\mathbf{B}$. This is also called the Riemann-Silberstein (RS) vector field [10]. Conversely, the QFT for light is constructed by quantizing the single-photon wave function. When the connections between the different formalisms are understood, it can be seen that there are at least two ways of interpreting the photon-wave-function theory. One interpretation is as a theory of quantum particles [17]. The other is to interpret PWF theory as an alternative means for describing states and dynamics of the quantum field in the case that we can restrict the description to a fixed subspace of the larger Fock space of photon numbers. Then, there is a clear relation between the PWF and the mode functions that appear in the state description of the quantum field. A subtlety arises when treating the Hilbert-space scalar product for PWFs. We present the Lorentz-invariant scalar product of the photon wave function, which is non-local in the coordinate representation. This form of scalar product can also be used to better interpret, understand, and utilize the Titulaer-Glauber wave-packet quantization formalism.

After showing how to switch from a photon viewpoint to a field viewpoint by quantization of the single-photon wave mechanics theory, we show how to switch in the opposite direction. That is, starting from conventional Dirac-Titulaer-Glauber QFT for light, we can extract the correct forms for photon wave functions and their equations of motion in coordinate space. This connects to similar treatments by Muthukrishnan et al, Lapaire and Sipe, and by Eberly et al. [3, 4, 18, 19]. This result shows how to incorporate interactions between light and matter in the photons viewpoint. A related example is that of Fini et al, who formulated the propagation of intense solitons in a Kerr third-order nonlinear optical medium in terms of the coordinate-space wave functions of many photons [20].

Single-photon spatial-temporal states can be used for a non-monochromatic wave-packet basis set in which to expand the electromagnetic field in QFT. As an example of this, we introduce a basis set of beam-like wave-packet modes with broad spectra. Such wave-packet modes are not orthogonal under the standard, local, coordinate-space scalar product, but are orthogonal under the non-local scalar product arising in photon wave mechanics. The closely related concept of the dual-mode basis is discussed in terms of orthogonalization of the T-G wave-packet modes, and leads directly to the non-local scalar product. We propose an experimental scheme to convert between wave-packet modes and their dual modes through a spectral-filtering scheme similar to a pulse shaper.

We extend the single-photon wave mechanics by developing two-photon and $n$-photon wave mechanics [21]. We show that there are deep connections between this extended photon-wave-mechanics theory and optical coherence theory—both quantum and classical. These connections are shown to be related to standard photodetection theory and to the standard wave-function collapse hypothesis. We also present a description of entanglement in the state of two photons, and the correct method for reduction of a pure two-photon wave-function state to a single-photon density matrix written in terms of single-photon wave functions. Here the non-local scalar product plays a crucial role in eliminating information about the traced-out photon.

## 2. From Monochromatic Modes to Wave-Packet Modes

We first develop the theory of photon wave packets (WP) in terms of QFT. The quantized electromagnetic theory developed by Dirac [12] starts from the classical Maxwell theory of the electromagnetic field, which is canonically quantized in terms of monochromatic modes. In free space, the electric and magnetic-induction field operators obey the Maxwell equations, in SI units,[1]

$$\partial_t \hat{\mathbf{E}}(\mathbf{x},t) = c^2 \nabla \times \hat{\mathbf{B}}(\mathbf{x},t), \quad \nabla \cdot \hat{\mathbf{E}}(\mathbf{x},t) = 0,$$
$$\partial_t \hat{\mathbf{B}}(\mathbf{x},t) = -\nabla \times \hat{\mathbf{E}}(\mathbf{x},t), \quad \nabla \cdot \hat{\mathbf{B}}(\mathbf{x},t) = 0. \tag{1}$$

The positive-frequency parts of the fields may be expanded using monochromatic modes as [12, 22-25]

$$\hat{\mathbf{E}}^{(+)}(\mathbf{x},t) = i \sum_\sigma \int \frac{d^3k}{(2\pi)^3} \left(\frac{\hbar c k}{2\varepsilon_0}\right)^{1/2} \hat{a}_{\mathbf{k},\sigma} \mathbf{u}_{\mathbf{k},\sigma}(\mathbf{x}) \exp(-i\omega_\mathbf{k} t), \tag{2}$$

$$\hat{\mathbf{B}}^{(+)}(\mathbf{x},t) = i \sum_\sigma \int \frac{d^3k}{(2\pi)^3} \left(\frac{\hbar c k}{2\varepsilon_0}\right)^{1/2} \hat{a}_{\mathbf{k},\sigma} \left(\frac{\mathbf{k}}{ck} \times \mathbf{u}_{\mathbf{k},\sigma}(\mathbf{x})\right) \exp(-i\omega_\mathbf{k} t), \tag{3}$$

where $\omega_\mathbf{k} = ck = c|\mathbf{k}|$, $c$ is the vacuum speed of light, and $\varepsilon_0$ is the permittivity of the vacuum. The monochromatic, orthonormal, plane-wave modes are

$$\mathbf{u}_{\mathbf{k},\sigma}(\mathbf{x}) = \mathbf{e}_{\mathbf{k},\sigma} \exp(i\mathbf{k}\cdot\mathbf{x}), \tag{4}$$

where the $\mathbf{e}_{\mathbf{k},\sigma}$ are unit polarization vectors. The sum is over the two mode-polarization indices $\sigma = \pm 1$. It turns out to be advantageous to assume circular polarization for modes (corresponding to positive and negative helicity), so we do this throughout this paper. For circular polarization

$$\mathbf{k} \times \mathbf{e}_{\mathbf{k},\sigma} = -i\sigma|\mathbf{k}|\mathbf{e}_{\mathbf{k},\sigma}, \tag{5}$$

although we prefer not to invoke this here, in order to keep (3) general. An advantage to using the plane-wave modes $\mathbf{u}_{\mathbf{k},\sigma}(\mathbf{x})$ is that they are orthogonal under the standard definition of the scalar product, which hereafter we call the *overlap integral*, and denote it by $(\ |\ )$. This is given by

$$\left(\mathbf{u}_{\mathbf{k},\sigma} | \mathbf{u}_{\mathbf{k}',\sigma'}\right) = \int \mathbf{u}_{\mathbf{k},\sigma}(\mathbf{x})^* \cdot \mathbf{u}_{\mathbf{k}',\sigma'}(\mathbf{x}) d^3x = (2\pi)^3 \delta^{(3)}(\mathbf{k}-\mathbf{k}')\delta_{\sigma,\sigma'}. \tag{6}$$

The Hermitian field operators are

$$\hat{\mathbf{E}}(\mathbf{x},t) = \hat{\mathbf{E}}^{(+)}(\mathbf{x},t) + \hat{\mathbf{E}}^{(-)}(\mathbf{x},t), \quad \hat{\mathbf{B}}(\mathbf{x},t) = \hat{\mathbf{B}}^{(+)}(\mathbf{x},t) + \hat{\mathbf{B}}^{(-)}(\mathbf{x},t), \tag{7}$$

where the negative-frequency parts are given by Hermitian conjugates of the positive-frequency operators, $\hat{\mathbf{E}}^{(-)}(\mathbf{x},t) = \left[\hat{\mathbf{E}}^{(+)}(\mathbf{x},t)\right]^\dagger$ and $\hat{\mathbf{B}}^{(-)}(\mathbf{x},t) = \left[\hat{\mathbf{B}}^{(+)}(\mathbf{x},t)\right]^\dagger$. The monochromatic annihilation and creation operators $\hat{a}_{\mathbf{k},\sigma}$ and $\hat{a}^\dagger_{\mathbf{k},\sigma}$ obey the bosonic commutation relations

$$\left[\hat{a}_{\mathbf{k},\sigma}, \hat{a}^\dagger_{\mathbf{k}',\sigma'}\right] = (2\pi)^3 \delta^{(3)}(\mathbf{k}-\mathbf{k}')\delta_{\sigma,\sigma'}. \tag{8}$$

Excitation-number operators are $\hat{n}_{\mathbf{k},\sigma} = \hat{a}^\dagger_{\mathbf{k},\sigma}\hat{a}_{\mathbf{k},\sigma}$. (We refrain from using the word *photon* here.) The electromagnetic-field Hamiltonian operator is expressed in terms of the annihilation and creation operators as

$$\hat{H} = \sum_\sigma \int \frac{d^3k}{(2\pi)^3} \hbar\omega_\mathbf{k} \hat{a}^\dagger_{\mathbf{k},\sigma}\hat{a}_{\mathbf{k},\sigma}, \tag{9}$$

---

[1] We "derive" the Maxwell equations in Section 4.

where we follow the common practice of neglecting the infinite vacuum energy term. The interaction of the quantized electromagnetic field with atomic systems can be introduced through an interaction term (usually the electric-dipole interaction in non-relativistic treatments) in the atom-field Hamiltonian.

The free-space field operators in (2) and (3) are expressed in terms of plane waves, which serve well for simple models. However, when localized space-time interactions are considered, such as spontaneous emission from an atom, [4, 11], the plane-wave description becomes inefficient. In such a case, Titulaer and Glauber (T-G) showed that one may expand the electric and magnetic fields in terms of non-orthogonal, non-monochromatic, spatial-temporal modes $\mathbf{v}_{j,\sigma}(\mathbf{x},t)$ [7]. Such modes were used to study transient Raman scattering [26], and have been recently discussed in terms of the transverse spatial modes of light [27]. These wave-packet modes are related to the orthogonal, monochromatic, plane-wave modes through the non-unitary transformation

$$\mathbf{v}_{j,\sigma}(\mathbf{x},t) = i\left(\frac{\hbar c}{2\varepsilon_0}\right)^{1/2} \int \frac{d^3k}{(2\pi)^3} \sqrt{k}\, U_j^{(\sigma)}(\mathbf{k}) \mathbf{u}_{\mathbf{k},\sigma}(\mathbf{x}) \exp(-i\omega_{\mathbf{k}} t), \qquad (10)$$

in which $U_j^{(\sigma)}(\mathbf{k})$ is a unitary transformation "matrix." Equation (4) shows that this relation is a Fourier transform, which can be inverted to give

$$U_j^{(\sigma)}(\mathbf{k})\mathbf{e}_{\mathbf{k},\sigma} = \left(\frac{2\varepsilon_0}{\hbar c}\right)^{1/2} \exp(i\omega_{\mathbf{k}} t) \frac{1}{i\sqrt{k}} \int d^3x\, \mathbf{v}_{j,\sigma}(\mathbf{x},t) \exp(-i\mathbf{k}\cdot\mathbf{x}). \qquad (11)$$

We call $\mathbf{v}_{j,\sigma}(\mathbf{x},t)$ the wave-packet (WP) modes. The "matrix" $U_j^{(\sigma)}(\mathbf{k})$ is unitary, that is, for fixed value of $\sigma$,

$$\sum_j U_j^{(\sigma)}(\mathbf{k}')^* U_j^{(\sigma)}(\mathbf{k}) = (2\pi)^3 \delta^{(3)}(\mathbf{k}'-\mathbf{k}), \qquad \int \frac{d^3k}{(2\pi)^3} U_j^{(\sigma)}(\mathbf{k}) U_{j'}^{(\sigma)}(\mathbf{k})^* = \delta_{jj'}. \qquad (12)$$

Nevertheless, the relation between the WP modes and the monochromatic modes is not unitary because of the $\sqrt{k}$ factor in (10). This turns out to be a crucial point.

The annihilation and creation operators are changed by the unitary transformation, leading to new annihilation and creation operators $\hat{b}_{j,\sigma}$ and $\hat{b}_{j,\sigma}^\dagger$ given by

$$\hat{b}_{j,\sigma} = \int \frac{d^3k}{(2\pi)^3} U_j^{(\sigma)}(\mathbf{k})^* \hat{a}_{\mathbf{k},\sigma}, \qquad (13)$$

which obey bosonic commutation relations $\left[\hat{b}_{j,\sigma},\hat{b}_{m,\rho}^\dagger\right] = \delta_{j,m}\delta_{\sigma,\rho}$. This means that one can construct states of definite excitation number $N$ in a particular WP mode by applying the creation operator to the vacuum state: $(\hat{b}_{j,\sigma}^\dagger)^N/\sqrt{N!}|\text{vacuum}\rangle = |N\rangle_{j,\sigma}$. The inverse of (13) is

$$\hat{a}_{\mathbf{k},\sigma} = \sum_j U_j^{(\sigma)}(\mathbf{k}) \hat{b}_{j,\sigma}. \qquad (14)$$

In terms of the WP modes, the positive-frequency parts of the electric and magnetic field operators are

$$\hat{\mathbf{E}}^{(+)}(\mathbf{x},t) = \sum_{j,\sigma} \hat{b}_{j,\sigma}\, \mathbf{v}_{j,\sigma}(\mathbf{x},t), \qquad (15)$$

$$\hat{\mathbf{B}}^{(+)}(\mathbf{x},t) = \sum_{j,\sigma} \hat{b}_{j,\sigma} \left(\frac{\mathbf{k}_j}{c|\mathbf{k}_j|} \times \mathbf{v}_{j,\sigma}(\mathbf{x},t)\right). \qquad (16)$$

The monochromatic plane-wave basis functions $\mathbf{u}_{\mathbf{k},\sigma}(\mathbf{x})$ are orthogonal under the standard definition of the scalar product, that is the overlap integral, (6). In contrast, the WP modes $\mathbf{v}_{j,\sigma}(\mathbf{x},t)$, do not generally form an orthogonal set under a scalar product defined by the overlap integral.[2] That is,

$$\left(\mathbf{v}_{j,\sigma}\middle|\mathbf{v}_{m,\sigma}\right) = \int \mathbf{v}_{j,\sigma}(\mathbf{x},t)^* \cdot \mathbf{v}_{m,\sigma}(\mathbf{x},t) d^3x = \frac{\hbar c}{2\varepsilon_0}\int \frac{d^3k}{(2\pi)^3} k\, U_j^{(\sigma)}(\mathbf{k})^* U_m^{(\sigma)}(\mathbf{k}) \neq \delta_{jm}. \quad (17)$$

The non-orthogonality arises from different weightings given to different frequency components by the $\sqrt{k}$ factor in (10). In fact, the WP modes are overcomplete, which might appear to be a disadvantage to using a WP expansion. However, as we next show, this may be overcome by defining a new scalar product for the wave-packet modes. This also leads to a crucial link between the WP modes and photon wave functions, treated in the next section.

Our extension of the T-G formalism rests on the definition of a new scalar product, under which the WP modes form an orthonormal set. To find the form of this scalar product, we introduce the well-known concept of a dual basis [28]. For every WP mode $\mathbf{v}_{j,\sigma}(\mathbf{x},t)$, we introduce a dual mode $\mathbf{v}_{j,\sigma}^D(\mathbf{x},t)$, such that

$$\left(\mathbf{v}_{j,\sigma}^D\middle|\mathbf{v}_{m,\rho}\right) = \delta_{jm}\delta_{\sigma\rho}. \quad (18)$$

In the present case, the dual modes are given by

$$\mathbf{v}_{j,\sigma}^D(\mathbf{x},t) = i\left(\frac{2\varepsilon_0}{\hbar c}\right)^{1/2} \int \frac{d^3k}{(2\pi)^3} \frac{1}{\sqrt{k}} U_j^{(\sigma)}(\mathbf{k}) \mathbf{u}_{\mathbf{k},\sigma}(\mathbf{x}) \exp(-i\omega_{\mathbf{k}}t). \quad (19)$$

This is nearly the same as (10), except that the $\sqrt{k}$ is in the denominator of the integrand and we have inverted the constant factor $(\hbar c/2\varepsilon_0)^{1/2}$.[3] The overlap integral between any dual mode and a WP mode is

$$\left(\mathbf{v}_{j,\sigma}^D\middle|\mathbf{v}_{m,\rho}\right) = \int \mathbf{v}_{j,\sigma}^D(\mathbf{x},t)^* \cdot \mathbf{v}_{m,\rho}(\mathbf{x},t) d^3x = \delta_{jm}\delta_{\sigma\rho}. \quad (20)$$

That is, the dual modes and the WP modes form a biorthogonal basis system under a scalar product defined by the overlap integral. Therefore, the projection of the field onto a mode $\mathbf{v}_{j,\sigma}(\mathbf{x},t)$ is accomplished by integration of the positive-frequency part of the field dotted with its complex-conjugate dual mode $\mathbf{v}_{j,\sigma}^{D*}(\mathbf{x},t)$, that is,

$$\int \mathbf{v}_{j,\sigma}^D(\mathbf{x},t)^* \cdot \hat{\mathbf{E}}^{(+)}(\mathbf{x},t) d^3x = \hat{b}_{j,\sigma}. \quad (21)$$

Each dual-mode can be expressed as an integral over the corresponding WP mode $\mathbf{v}_{j,\sigma}(\mathbf{x},t)$. Inserting (11) into (19) gives

$$\mathbf{v}_{j,\sigma}^D(\mathbf{x},t) = \int \mathbf{v}_{j,\sigma}(\mathbf{x}',t) K(\mathbf{x}-\mathbf{x}') d^3x', \quad (22)$$

where the kernel is (see Appendix A)

$$K(\mathbf{x}) = \frac{2\varepsilon_0}{\hbar c}\int \frac{d^3k}{(2\pi)^3} \frac{\exp(i\mathbf{k}\cdot\mathbf{x})}{k} = \frac{2\varepsilon_0}{\hbar c}\frac{1}{2\pi^2}\frac{1}{|\mathbf{x}|^2}. \quad (23)$$

Thus, we can rewrite (20) as

$$\left(\mathbf{v}_{j,\sigma}^D\middle|\mathbf{v}_{m,\rho}\right) = \int d^3x \int d^3x'\, \mathbf{v}_{j,\sigma}(\mathbf{x}',t)^* \cdot \mathbf{v}_{m,\rho}(\mathbf{x},t) K(\mathbf{x}-\mathbf{x}') = \delta_{jm}\delta_{\sigma\rho}. \quad (24)$$

This shows that we can define a new form of scalar product for the WP modes, under which the WP modes are orthogonal. We denote this new scalar product by $(\ \|\ )$, where

$$\left(\mathbf{v}_{j,\sigma}\|\mathbf{v}_{m,\rho}\right) \equiv \left(\mathbf{v}_{j,\sigma}^{D}\big|\mathbf{v}_{m,\rho}\right) = \frac{\varepsilon_0}{\hbar c \pi^2} \int d^3x \int d^3x' \frac{\mathbf{v}_{j,\sigma}(\mathbf{x},t)^* \cdot \mathbf{v}_{m,\rho}(\mathbf{x}',t)}{|\mathbf{x}-\mathbf{x}'|^2} = \delta_{jm}\delta_{\sigma\rho}. \qquad (25)$$

---

[2] The wave-packet modes are also non-orthogonal under the integral $\iint d^3x\, dt$.

[3] Note that the dual modes are proportional to the wave-packet modes used for expanding the vector potential. This provides a link to the formalism of Hawton and Melde [29].

Such an integral is called *non-local* because the mode value at one point is multiplied by values of the other mode at every spatial point. Equation (25) is just another way to represent (24), but it has important meaning when discussing photon wave functions in the next sections. In fact, the same form of scalar product was first introduced precisely in the context of photon wave functions [30]. The present paper is, to our knowledge, the first to apply this form of scalar product in the context of WP field quantization.

An important state of light is that in which a single excitation occurs in a given spatial-temporal-localized packet. An example is the deterministic generation of a single photon from an atom in a cavity-QED system [2]. If the packet is dispersed spectrally by a prism and detected by an array of photon counters, only one counter will click, although which one clicks will be random. Such a state is expressed as

$$|1\rangle_{j,\sigma} = \hat{b}_{j,\sigma}^\dagger |\text{vacuum}\rangle = \int \frac{d^3k}{(2\pi)^3} U_j^{(\sigma)}(\mathbf{k}) |1\rangle_{\mathbf{k},\sigma}, \qquad (26)$$

where $|1\rangle_{\mathbf{k},\sigma}$ is a state with a single excitation having particular monochromatic wave vector-polarization state labelled by the pair $(\mathbf{k},\sigma)$. We see that the function $U_j^{(\sigma)}(\mathbf{k})$ fully specifies the state.

The simplest example one might construct to illustrate the use of the WP modes and the new scalar product (25), is a complete set of modes representing propagation in one dimension. This case runs into subtleties, because of the continuous nature of the $\mathbf{k}$ variable. These subtleties are familiar for non-normalizable modes in the continuum limit, such as infinite plane waves or Bessel beams. We defer the details of this discussion until after we have derived the photon-wave-function formalism.

## 3. Deriving the single-photon wave equation from Einstein kinematics

In the previous section we reviewed the standard Dirac quantum theory of electromagnetism [12], in which the classical Maxwell fields are raised to the status of operators acting on a Hilbert space— the state space of the electromagnetic field—but still obey the Maxwell equations. This is in contrast to the traditional quantum treatment of the electron, in which the single-particle wave function is first introduced to describe the relativistic evolution of one electron [13, 14]. This is then followed by quantization of the relativistic Dirac equation for a single particle, by elevating the wave function to the status of an operator. It is also common to omit the single-particle description, and develop the full quantum field theory from the beginning [31-34]. In this section we aim to show that the traditional particle-like approach may also be applied to the case of the photon.

Here we temporarily "forget" all that we discussed in Section 2, and begin by paralleling Dirac's approach to determining the wave equation for the single electron, a spin-1/2 particle, from Einstein kinematics. By this straightforward route, we arrive at the equation of motion for the wave function of the zero-mass, spin-1 photon, which is the same as found by other authors [8-11]. We discuss the normalization and scalar product of the photon wave function, noting that the scalar product must be a non-local integral in coordinate space of the form in (25) to ensure a local energy density in coordinate space. Finally, we quantize this single-particle theory in much the same way the Dirac theory of the electron is quantized in terms of creation and annihilation operators. By determining the energy eigenstates of the photon energy operator, we discover the connection between the quantized single-photon theory and the monochromatic-mode Dirac theory of electromagnetism.

We start from the Einstein energy-momentum-mass relationship

$$E = \left(c^2|\mathbf{p}|^2 + m^2c^4\right)^{1/2}, \qquad (27)$$

which Dirac used to derive his wave equation for the electron [13]. For the photon, the mass is $m=0$, so the energy is given in terms of the momentum,

$$E = c\sqrt{\mathbf{p}\cdot\mathbf{p}}. \qquad (28)$$

Our rationale in this part of the derivation [35] is closest to that of Sipe [11], who also started from the particle-like equation (28), whereas Bialynicki-Birula [8-10] started from the Maxwell equations describing light as a wave. Note that the number of wave-function components for a spin-$j$ particle, $j = 0, 1/2, 1, 3/2, ...$, is given by $n = (2j+1)$, which arises from the general treatment of rotations in three dimensions [17]. Assuming that the photon is a spin-1 particle, from empirical evidence, this leads to a three-component wave function for a photon of definite helicity.

Starting in the energy-momentum representation, where a momentum-space wave function can be well defined, we introduce a three-component (vector) momentum-space wave function $\tilde{\psi}(\mathbf{p})$. As for any vector field, $\tilde{\psi}(\mathbf{p})$ can be separated into transverse and longitudinal parts [25],

$$\tilde{\psi}(\mathbf{p}) = \tilde{\psi}^{(T)}(\mathbf{p}) + \tilde{\psi}^{(L)}(\mathbf{p}), \tag{29}$$

obeying

$$\mathbf{p} \cdot \tilde{\psi}^{(T)}(\mathbf{p}) = 0, \quad \mathbf{p} \times \tilde{\psi}^{(L)}(\mathbf{p}) = 0. \tag{30}$$

We follow the standard approach in quantum theory and assume that (28) is the dispersion relation for some first-order-in-time wave equation, which we wish to determine. To find it [35], we multiply (28) by the wave function $\tilde{\psi}(\mathbf{p})$ to give

$$E\tilde{\psi}(\mathbf{p}) = c\sqrt{\mathbf{p} \cdot \mathbf{p}}\, \tilde{\psi}(\mathbf{p}). \tag{31}$$

Substituting (29) into (31) shows that the transverse and longitudinal parts evolve independently in free space. We then make use of the vector identity

$$\mathbf{p} \times \mathbf{p} \times \tilde{\psi}^{(T)}(\mathbf{p}) = -\mathbf{p} \cdot \mathbf{p}\, \tilde{\psi}^{(T)} + \mathbf{p}\left(\mathbf{p} \cdot \tilde{\psi}^{(T)}\right) = -\mathbf{p} \cdot \mathbf{p}\, \tilde{\psi}^{(T)} \tag{32}$$

to write the equation for the transverse part as[4]

$$E\tilde{\psi}^{(T)}(\mathbf{p}) = \pm ic\mathbf{p} \times \tilde{\psi}^{(T)}(\mathbf{p}). \tag{33}$$

We now drop the *T* superscript, and interpret this equation as an energy-eigenvalue equation,

$$H\tilde{\psi}_\sigma(\mathbf{p}) = E\tilde{\psi}_\sigma(\mathbf{p}), \tag{34}$$

where we have introduced the index $\sigma = \pm 1$, which we will see corresponds to the helicity of the photon. The Hermitian Hamiltonian operator is defined as

$$H = ic\sigma\mathbf{p} \times. \tag{35}$$

Note that $H$ also depends on the value of $\sigma$ when treating the two helicities separately.

This equation can be put into a form that more closely resembles the Dirac equation with spin dependence [8, 11]. Note the following feature of the spin-1 matrices and vector cross product, $\mathbf{a} \times \mathbf{b} = -i(\mathbf{a} \cdot \mathbf{s})\mathbf{b}$, where $\mathbf{a}$ and $\mathbf{b}$ are ordinary three-component vectors, and $\mathbf{s} = (s_x, s_y, s_z)$ is a three-component vector composed of the three spin-1 matrices (generators of rotations for spin-1 particles)

$$s_x = \begin{bmatrix} 0 & 0 & 0 \\ 0 & 0 & -i \\ 0 & i & 0 \end{bmatrix}, \quad s_y = \begin{bmatrix} 0 & 0 & i \\ 0 & 0 & 0 \\ -i & 0 & 0 \end{bmatrix}, \quad s_z = \begin{bmatrix} 0 & -i & 0 \\ i & 0 & 0 \\ 0 & 0 & 0 \end{bmatrix}. \tag{36}$$

This leads to the following form of the photon Hamiltonian in terms of the spin matrices

$$H = ic\sigma\mathbf{p} \times = c\sigma(\mathbf{s} \cdot \mathbf{p}), \tag{37}$$

and the corresponding momentum-space wave equation

$$E\tilde{\psi}_\sigma(\mathbf{p}) = c\sigma(\mathbf{s} \cdot \mathbf{p})\tilde{\psi}_\sigma(\mathbf{p}) = c|\mathbf{p}|\tilde{\psi}_\sigma(\mathbf{p}). \tag{38}$$

The helicity dependence is now explicitly present, as can be seen by noting that the helicity operator, the projection of the spin onto the direction of propagation, is

$$\hat{h} = \frac{\mathbf{p}}{|\mathbf{p}|} \cdot \mathbf{s}. \tag{39}$$

---

[4] The requirement to find a first-order-in-time wave equation has forced us to discard the longitudinal part of the field, at least for evolution in free space. When interactions with charged particles are considered, the longitudinal, near-field part may come back [36].

In general, one must treat both helicities on equal footing, which can be done by creating a six-component, spinor wave function [8-10]. However, in free space the helicities do not mix, so in this case we can treat each helicity independently.

The task is to transform (38) into a coordinate-space wave equation in a way that clearly represents the known physics of photons. The interpretation of the momentum-space photon wave function $\tilde{\psi}_\sigma(\mathbf{p})$, introduced above, must be addressed prior to transforming to coordinate space. The momentum-space wave function $\tilde{\psi}_\sigma(\mathbf{p})$ is typically interpreted as the probability amplitude in momentum space. This means that $|\tilde{\psi}_\sigma(\mathbf{p})|^2 d^3p (2\pi\hbar)^{-3}$ gives the probability of finding a photon (anywhere in space) with helicity $\sigma$, and momentum in a momentum-space volume $d^3p$ about $\mathbf{p}$. The proper normalization for normalizable momentum-space wave functions is thus [11, 17][5]

$$(\psi\|\psi) = \sum_\sigma \int \frac{d^3p}{(2\pi\hbar)^3} \tilde{\psi}_\sigma(\mathbf{p})^\dagger \tilde{\psi}_\sigma(\mathbf{p}) = 1. \tag{40}$$

We use the notation $(\psi\|\psi)$ for the norm to remind us that this might not correspond to an overlap integral of the form (6) when written in coordinate space. (The dagger indicates complex conjugation as well as vector transposition; therefore we need no explicit "dot" operator as we used in the previous section for mode functions.) For eigenstates of momentum and helicity, $\tilde{\psi}_{\mathbf{p},\sigma}$, which are not normalizable, (40) is replaced by

$$(\tilde{\psi}_{\mathbf{p}',\sigma'} \| \tilde{\psi}_{\mathbf{p}'',\sigma''}) = \int \frac{d^3p}{(2\pi\hbar)^3} \tilde{\psi}_{\mathbf{p}',\sigma'}(\mathbf{p})^\dagger \tilde{\psi}_{\mathbf{p}'',\sigma''}(\mathbf{p}) = \delta_{\sigma'\sigma''} \delta^{(3)}(\mathbf{p}' - \mathbf{p}''), \tag{41}$$

from which we infer

$$\tilde{\psi}_{\mathbf{p}',\sigma'}(\mathbf{p}) = (2\pi\hbar)^{3/2} \mathbf{e}_{\mathbf{p}',\sigma'} \delta^{(3)}(\mathbf{p} - \mathbf{p}'). \tag{42}$$

In standard non-relativistic quantum mechanics for massive particles, momentum-space wave functions and coordinate-space wave functions, which are interpreted as probability amplitudes in coordinate space, are related by Fourier transforms. However, it is well known that photons, being inherently relativistic particles, are not localizable and have no well-defined coordinate-space eigenstates [15]. Thus one may *not* interpret the Fourier transform of $\tilde{\psi}_\sigma(\mathbf{p})$, given by

$$\psi_\sigma^{LP}(\mathbf{x},t) = \int \frac{d^3p}{(2\pi\hbar)^3} \exp\left[i(\mathbf{p}\cdot\mathbf{x} - c|\mathbf{p}|t)/\hbar\right] \tilde{\psi}_\sigma(\mathbf{p}), \tag{43}$$

as a coordinate-space wave function.[6] Nonetheless, this has been done in the past and interpreted, albeit not very usefully, as the photon wave function [37-40]. We have denoted this "wave function" with the superscript *LP* for Landau and Peierls, the first to propose this form as a candidate for a photon wave function in coordinate space. There are several reasons for not choosing this function as the true single-photon wave function, including that this function is non-locally connected[7] to the classical electromagnetic field [10, 17], it does not transform as any geometric object under Lorentz transformations, and it behaves non-causally when considering spontaneous emission by an excited atom [10, 11, 41]. However, in the quasi-monochromatic approximation this theory can be used to simplify some calculations [4].

---

[5] This is Sipe's convention [11]. One can adopt different conventions for this normalization, such as $\sum_\sigma \int d^3p |\mathbf{p}|^{-1} (2\pi\hbar)^{-3} \tilde{\psi}_\sigma^{BB}(\mathbf{p})^\dagger \tilde{\psi}_\sigma^{BB}(\mathbf{p}) = 1$, as in Appendix B and in [8-10]. This changes the definition of the wave function to the form used by Bialynicki-Birula: $\tilde{\psi}_\sigma^{BB}(\mathbf{p}) = \sqrt{c|\mathbf{p}|}\, \tilde{\psi}_\sigma(\mathbf{p})$.

[6] Equation (43) is valid for non-relativistic electrons because $\langle \mathbf{x}|\mathbf{p}\rangle = \exp(-i\mathbf{p}\cdot\mathbf{x}/\hbar)$ provides a relation between momentum eigenstates and position eigenstates, whereas the latter do not exist for photons. This is only approximately valid for a non-relativistic electron where its speed is much less than $c$.

To obtain the coordinate-space representation of (38), we weight the Fourier transformation with a function $f(|\mathbf{p}|)$ of the magnitude of the momentum (equivalently, the energy) [11]. This leads to the following wave function form

$$\psi_\sigma(\mathbf{x},t) = \int \frac{d^3p}{(2\pi\hbar)^3} \exp\left[i(\mathbf{p}\cdot\mathbf{x} - c|\mathbf{p}|t)/\hbar\right] f(|\mathbf{p}|)\tilde{\psi}_\sigma(\mathbf{p}), \qquad (44)$$

where the function $f(|\mathbf{p}|)$ is yet to be determined. Using (38), the wave equation obeyed by any function of the form in (44) is found to be[8]

$$i\hbar\partial_t \psi_\sigma(\mathbf{x},t) = \hbar c\sigma \nabla \times \psi_\sigma(\mathbf{x},t), \qquad (45)$$

or equivalently,

$$i\hbar\partial_t \psi_\sigma(\mathbf{x},t) = -i\hbar c\sigma(\mathbf{s}\cdot\nabla)\psi_\sigma(\mathbf{x},t), \qquad (46)$$

along with the implied zero divergence. This is the correct Schrödinger-like equation for the coordinate-space wave function of a single photon with a fixed helicity. We call this the *photon wave equation*. However, it provides no useful physics unless we can judicially choose the form of the weighting function $f(|\mathbf{p}|)$, which determines the interpretation of the coordinate-space wave function.

A rigorous way to ascertain the weighting function $f(|\mathbf{p}|)$ is to consider the form of the energy expectation value in coordinate space. Consistent with standard atomic physics and quantum optics, we choose to require that there exist a local energy density describing light. We thus require the energy expectation value to be an integral of a localized quantity in coordinate space. In momentum space, the energy expectation value is simply given by

$$\langle \hat{H} \rangle = \sum_\sigma \int \frac{d^3p}{(2\pi\hbar)^3} \tilde{\psi}_\sigma(\mathbf{p})^\dagger \hat{H}\tilde{\psi}_\sigma(\mathbf{p}) = \sum_\sigma \int \frac{d^3p}{(2\pi\hbar)^3} c|\mathbf{p}| \tilde{\psi}_\sigma(\mathbf{p})^\dagger \tilde{\psi}_\sigma(\mathbf{p}). \qquad (47)$$

To determine the form of the energy expectation value in coordinate space we first invert the weighted Fourier-transform (44). This gives the momentum-space wave function in terms of the coordinate-space wave function

$$\tilde{\psi}_\sigma(\mathbf{p}) = \frac{1}{f(|\mathbf{p}|)} \int d^3x \exp\left[-i(\mathbf{p}\cdot\mathbf{x} - c|\mathbf{p}|t)/\hbar\right] \psi_\sigma(\mathbf{x},t). \qquad (48)$$

When transformed to coordinate space the energy expectation value, (47), thus becomes

$$\langle \hat{H} \rangle = \sum_\sigma \int d^3x \int d^3x' \psi_\sigma(\mathbf{x},t)^\dagger \psi_\sigma(\mathbf{x}',t) \int \frac{d^3p}{(2\pi\hbar)^3} \frac{c|\mathbf{p}|}{|f(|\mathbf{p}|)|^2} \exp\left[i\mathbf{p}\cdot(\mathbf{x}-\mathbf{x}')/\hbar\right]. \qquad (49)$$

In order for this expression of the energy expectation value to be local (that is, an integral over a local energy density), the square modulus of the weight function must cancel out the $c|\mathbf{p}|$ in (49), yielding a Dirac delta function. This gives a local expression for the coordinate-space energy expectation value

$$\langle \hat{H} \rangle = \sum_\sigma \int d^3x \, \psi_\sigma(\mathbf{x},t)^\dagger \psi_\sigma(\mathbf{x},t). \qquad (50)$$

Thus we see that the weight function is given by

$$f(|\mathbf{p}|) = \sqrt{c|\mathbf{p}|} = \sqrt{E}. \tag{51}$$

---

[7] By non-locally connected we mean that two functions $f$ and $g$ are related by $f(x) = \int g(x')J(x,x')dx'$, for some non-delta-function kernel function $J(x,x')$.

[8] Planck's constant cancels from both sides of this equation of motion, although it reappears when one calculates energy-related quantities.

As stated above, the single-photon wave function in momentum space $\tilde{\psi}_\sigma(\mathbf{p})$ is the probability amplitude for finding the photon with polarization (helicity) $\sigma$ and momentum $\mathbf{p}$. The probabilistic interpretation of quantum mechanics requires a definition of a scalar product between two different states $\phi$ and $\psi$, denoted by $(\phi\|\psi)$, which is used in calculating transition probabilities. The Born rule states that the modulus squared of the scalar product of two normalized wave functions $|(\phi\|\psi)|^2$ is to be interpreted as the probability of observing a photon "in" state $\phi$ when it is known to be described by state $\psi$. The probability is a real, dimensionless number (a scalar), and thus must be invariant under any Poincaré transformation (Lorentz boost plus translation and rotation). From (50) we see that the overlap integral (standard scalar product) (6), cannot work since it gives the energy expectation value, which is not and should not be Lorentz invariant. Equations (40) and (41) imply that the scalar product in momentum space is of the familiar form

$$(\phi\|\psi) \equiv \sum_\sigma \int \frac{d^3p}{(2\pi\hbar)^3} \tilde{\phi}_\sigma(\mathbf{p})^\dagger \tilde{\psi}_\sigma(\mathbf{p}). \tag{52}$$

To determine the form of the coordinate-space scalar product corresponding to the momentum-space scalar product in (52), we use the inverted form of the weighted Fourier-transform (48). Inserting this expression of the momentum-space wave function in the scalar product, (52), we obtain the coordinate-space scalar product

$$(\phi\|\psi) = \int d^3x \int d^3x' \phi(\mathbf{x},t)^\dagger \psi(\mathbf{x}',t) \int \frac{d^3p}{(2\pi\hbar)^3} \frac{1}{|f(|\mathbf{p}|)|^2} \exp[i\mathbf{p}\cdot(\mathbf{x}-\mathbf{x}')/\hbar]$$

$$= \int d^3x \int d^3x' \phi(\mathbf{x},t)^\dagger \psi(\mathbf{x}',t) G(\mathbf{x}-\mathbf{x}'), \tag{53}$$

where the kernel function $G(\mathbf{x}-\mathbf{x}')$ is given by

$$G(\mathbf{x}-\mathbf{x}') = \int \frac{d^3p}{(2\pi\hbar)^3} \frac{1}{|f(|\mathbf{p}|)|^2} \exp[i\mathbf{p}\cdot(\mathbf{x}-\mathbf{x}')/\hbar]$$

$$= \int \frac{d^3p}{(2\pi\hbar)^3} \frac{1}{c|\mathbf{p}|} \exp[i\mathbf{p}\cdot(\mathbf{x}-\mathbf{x}')/\hbar] = \frac{1}{2\pi^2\hbar c} \frac{1}{|\mathbf{x}-\mathbf{x}'|^2}. \tag{54}$$

Here the wave functions are defined with two components to include the state for both helicities,

$$\psi(\mathbf{x},t) \equiv \begin{bmatrix} \psi_{+1}(\mathbf{x},t) \\ \psi_{-1}(\mathbf{x},t) \end{bmatrix}, \tag{55}$$

that is, $\psi(\mathbf{x},t)$ is a six-component object [10]. Note that the choice of weight function, (51), ensures that the kernel function $G(\mathbf{x}-\mathbf{x}')$ in (53) transforms as a Lorentz-scalar function (see Appendix C). The Lorentz-invariant scalar product is thus a non-local integral in coordinate space, given by

$$(\phi\|\psi) = \frac{1}{2\pi^2\hbar c} \int d^3x \int d^3x' \frac{\phi(\mathbf{x},t)^\dagger \psi(\mathbf{x}',t)}{|\mathbf{x}-\mathbf{x}'|^2}. \tag{56}$$

For a normalizable state the norm in coordinate space is then

$$(\psi\|\psi) = \frac{1}{2\pi^2\hbar c} \int d^3x \int d^3x' \frac{\psi(\mathbf{x},t)^\dagger \psi(\mathbf{x}',t)}{|\mathbf{x}-\mathbf{x}'|^2} = 1. \tag{57}$$

The Lorentz invariance of the norm corresponds to the physical fact that one does not "create" or "destroy" photons simply by viewing a situation from a different inertial reference frame.

The scalar product (53) and norm (57) are non-local integrals, which is not surprising in light of the well-known fact that there exists no such quantity as photon number density. In fact, it has been claimed that one of the few localizable quantities that can be associated with a photon is its energy [10, 11]. For the photon, the momentum, angular momentum, and the moment of energy are also local quantities [10]. It has been shown that for massless particles with spin greater than one, even the energy is non-localizable [42].

For comparison, recall that in non-relativistic (Schrödinger) *electron* wave mechanics, the overlap integral (6) is normalized to one. For electrons, one can interpret $m_e |\psi(\mathbf{x},t)|^2$ (or $e|\psi(\mathbf{x},t)|^2$) as an average local mass (or charge) density. Since the electron rest mass $m_e$ (or charge) is a constant, this function can be normalized to unity, as is usually done. For photons, there is no mass density (nor a charge density), and we find the energy density to be the appropriate local concept. Since the photon energy is not a constant, it makes no sense from this viewpoint to look for a local number density that is normalized to one.

We conclude, then, that a single-photon state of light can be described as if a single particle is present, whose state is described by the general wave function,

$$\psi(\mathbf{x},t) = \sum_{j,\sigma} B_{j,\sigma} \psi_{j,\sigma}(\mathbf{x},t), \tag{58}$$

where the $B_{j,\sigma}$ are expansion amplitudes and $\{\psi_{j,\sigma}(\mathbf{x},t)\}$ is a complete set of states that are orthonormal with respect to the non-local scalar product defined in (52), that is,

$$\left(\psi_{j,\sigma} \| \psi_{k,\rho}\right) = \frac{1}{2\pi^2 \hbar c} \int d^3x \int d^3x' \frac{\psi_{j,\sigma}(\mathbf{x},t)^\dagger \psi_{k,\rho}(\mathbf{x}',t)}{|\mathbf{x}-\mathbf{x}'|^2} = \delta_{jk}\delta_{\sigma\rho}. \tag{59}$$

Notice that, following [10], we use the word *states* (rather than *modes*) here for the functions $\psi_{j,\sigma}(\mathbf{x},t)$, consistent with the usage for electrons.

The matrix element of any operator in a given basis $\{\psi_j(\mathbf{x},t)\}$ (suppressing the helicity label) may be conveniently expressed in a form analogous to standard Dirac bra-ket notation as

$$\left(\psi_j \| \hat{O} | \psi_k\right) = \frac{1}{2\pi^2 \hbar c} \int d^3x \int d^3x' \frac{\psi_j(\mathbf{x},t)^\dagger}{|\mathbf{x}-\mathbf{x}'|^2} \hat{O} \psi_k(\mathbf{x}',t). \tag{60}$$

Our notation $(\psi \| \hat{O} | \psi)$ emphasizes that the operator $\hat{O}$ acts on the ket to its right, $|\psi)$, while the double-lined bra $(\psi\|$ indicates the linear functional

$$\left(\psi_j \right\| \bullet = \frac{1}{2\pi^2 \hbar c} \int d^3x \int d^3x' \frac{1}{|\mathbf{x}-\mathbf{x}'|^2} \psi_j(\mathbf{x}',t)^\dagger \bullet, \tag{61}$$

where $\bullet$ indicates any function of $\mathbf{x}$.

In many of the previous treatments of the photon wave function [37-40, 43] either the non-local wave function (43), or the incorrect (non-Lorentz-invariant) scalar product, similar to (6), was used. The *LP* wave function, (43) does not have a local interaction with charged particles [10]. Thus the vanishing of the *LP* wave function at a given point does not imply absence of interaction with a charge at that point. Although the proper energy-density wave function is discussed in [43], the proper scalar product is not utilized. This approach [43] is nearly identical to the course-grained photon number density approach of Mandel [44], in which one may approximately localize a photon in a volume larger than its cubic wavelength. This approach is also similar to that taken by Fedorov et al to describe spontaneous emission of a photon from an atom and the resulting atom-photon entanglement [4]. Here the authors are careful to note that their method is valid only for quasi-monochromatic light. Recent work using a Lagrangian approach, rather than a Hamiltonian approach, has shown to be useful for deriving conservation laws for single-photon states [45] and introducing interactions with charged particles [36].

## 4. Quantization of the single-photon wave function

We see a remarkable confluence between the form of the scalar product necessary for wave-packet modes to be orthogonal and the form of the scalar product found for photon wave functions (states). That is, (25) and (59) are identical in form. This recognition shows that the non-local scalar product occurs naturally in two contexts. This also guides us to an easy route for quantizing the single-photon theory to build up a quantum field theory (QFT) for describing states with more than one photon present. This was described briefly in [10, 43, 46], and we go beyond that discussion.

To construct a QFT of light, we raise the six-component photon wave function $\psi(\mathbf{x},t)$ in (58) to the status of a field operator. Its equation of motion follows from the same algebra we applied to the PWF introduced earlier in (31). We expand the wave function in a complete set of modes $\{\psi_{j,\sigma}(\mathbf{x},t)\}$, each obeying (45), which are orthonormal with respect to the non-local scalar product in (59), and replace the expansion amplitudes by annihilation and creation operators, $\hat{b}_{j,\sigma}$ and $\hat{b}^\dagger_{j,\sigma}$. The photon field operator is then defined as

$$\hat{\Psi}(\mathbf{x},t) = \sum_\sigma \hat{\Psi}_\sigma(\mathbf{x},t) = \sum_{j,\sigma} \hat{b}_{j,\sigma}\, \psi_{j,\sigma}(\mathbf{x},t) + H.c., \tag{62}$$

where *H.c.* stands for Hermitian conjugate. The canonical boson commutation relations for the $\hat{b}_{j,\sigma}$ and $\hat{b}^\dagger_{j,\sigma}$ operators are

$$\left[\hat{b}_{j,\sigma}, \hat{b}^\dagger_{k,\rho}\right] = \delta_{jk}\delta_{\sigma\rho}. \tag{63}$$

Each helicity component of the field operator obeys the equation of motion

$$i\hbar\partial_t \hat{\Psi}_\sigma(\mathbf{x},t) = \hbar c \sigma \nabla \times \hat{\Psi}_\sigma(\mathbf{x},t). \tag{64}$$

We next show that this theory, (62) – (64), is equivalent to the extended Titulaer-Glauber (T-G) version of QFT we discussed earlier (compare to (15) and (16)), with the single-photon basis states $\psi_{j,\sigma}(\mathbf{x},t)$ playing the same role as played by the wave-packet modes $\mathbf{v}_{j,\sigma}(\mathbf{x},t)$ in the T-G theory. This is a main result of the present paper. To make the correspondence clear, first note that if we break each basis state into real and imaginary parts,

$$\psi_{j,\sigma}(\mathbf{x},t) = \psi_j^R(\mathbf{x},t) + i\sigma \psi_j^I(\mathbf{x},t), \tag{65}$$

they obey the coupled wave equations

$$\partial_t \psi_j^R(\mathbf{x},t) = c\nabla \times \psi_j^I(\mathbf{x},t), \qquad \partial_t \psi_j^I(\mathbf{x},t) = -c\nabla \times \psi_j^R(\mathbf{x},t), \tag{66}$$

as can be seen by separating the complex wave equation (45) into two real equations. Along with the implicit zero-divergence conditions,

$$\nabla \cdot \psi_j^R(\mathbf{x},t) = \nabla \cdot \psi_j^I(\mathbf{x},t) = 0, \tag{67}$$

these equations, (66) and (67), are identical in form to the free-space Maxwell equations, with the real part of the photon wave function playing the role of the electric field and the imaginary part of the photon wave function playing the role of the magnetic-induction field (multiplied by *c*). This might be seen as remarkable, since it implies that Maxwell in 1864 found the relativistic quantum wave equation for single photons, but, of course, did not realize it.

If we analogously break the photon field operator $\hat{\Psi}_\sigma(\mathbf{x},t)$ into "real" and "imaginary" Hermitian parts,

$$\hat{\Psi}_\sigma(\mathbf{x},t) = \left(\frac{\varepsilon_0}{2}\right)^{1/2}\left[\hat{\mathbf{E}}(\mathbf{x},t) + i\sigma c \hat{\mathbf{B}}(\mathbf{x},t)\right], \tag{68}$$

then, from (64), (68), and the zero-divergence condition, we derive equations identical to (1), that is, the Maxwell equations. So, the real and imaginary parts of the operator $\hat{\Psi}_\sigma(\mathbf{x},t)$ equal the quantized electric and magnetic field operators in the standard QFT. The complex sum of electric and magnetic fields in (68) is called the Riemann-Silberstein (RS) vector, and has many useful properties [10, 46, 47].

A difference between the present field quantization (62) and the familiar Dirac QFT is that in (62) we use non-monochromatic, wave-packet modes as the space-time basis, whereas Dirac used monochromatic modes. Since we have already seen that the single-photon basis states $\psi_{j,\sigma}(\mathbf{x},t)$ play the same role as played by the wave-packet modes $\mathbf{v}_{j,\sigma}(\mathbf{x},t)$ in the T-G theory, we conclude that we have derived the latter directly, by starting from the single-photon dynamics represented by (28), and quantizing this single-photon theory. This is a novel way to derive QFT for light.

Before we study further ramifications, we wish to show how to convert the present T-G QFT (62) into the more familiar Dirac version, which uses monochromatic modes. The mode functions $\{\psi_{j,\sigma}(\mathbf{x},t)\}$, are in general non-monochromatic, that is they are not energy eigenstates. They are orthonormal under the non-local scalar product defined in (59). To find the relationship between the monochromatic Dirac and T-G theories, we must find the energy eigenstates of the derived T-G theory. Here we consider a fixed value of helicity $\sigma$ and suppress it in the notation. One may always treat the two helicity sets separately when considering free-space propagation. To find the energy eigenstates, we define an energy matrix in a particular basis $\{\psi_j(\mathbf{x},t)\}$, with non-zero off-diagonal elements $H_{jk}$, as

$$H_{jk} = (\psi_j \| \hat{H} | \psi_k) = \frac{1}{2\pi^2 \hbar c} \int d^3x \int d^3x' \frac{1}{|\mathbf{x}-\mathbf{x}'|^2} \psi_j(\mathbf{x}',t)^\dagger \hat{H} \psi_k(\mathbf{x},t)$$

$$= \frac{1}{2\pi^2 \hbar c} \int d^3x \int d^3x' \frac{1}{|\mathbf{x}-\mathbf{x}'|^2} \psi_j(\mathbf{x}',t)^\dagger (-i\hbar c \mathbf{s} \cdot \nabla) \psi_k(\mathbf{x},t) \qquad (69)$$

$$= \int d^3x \, \psi_j(\mathbf{x},t)^\dagger \psi_k(\mathbf{x},t) = (\psi_j | \psi_k).$$

Here we have made connection with the earlier notation in (6) for the overlap integral on the last line of (69). This form of the energy matrix is most easily verified by transforming to momentum space. The last line in (69) again indicates that it is the mean energy density $\psi_j(\mathbf{x},t)^\dagger \psi_j(\mathbf{x},t)$, and not photon number, that is localized in coordinate space.

Because the Hamiltonian is Hermitian, we can diagonalize this matrix using a unitary matrix $U_{jk}$. This transforms the modes $\{\psi_j(\mathbf{x},t)\}$ to another set of modes $\{\tilde{\phi}_k(\mathbf{x},t)\}$ given by

$$\tilde{\phi}_k(\mathbf{x},t) = \sum_j U_{jk} \psi_j(\mathbf{x},t), \qquad (70)$$

which also are orthonormal under the scalar product defined in (59). The operators are also transformed, leading to new creation and annihilation operators,

$$\hat{a}_k = \sum_j U_{jk}^\dagger \hat{b}_j, \qquad (71)$$

and

$$\hat{a}_k^\dagger = \sum_j U_{jk} \hat{b}_j^\dagger, \qquad (72)$$

which also obey the standard, bosonic commutation relations

$$\left[\hat{a}_j, \hat{a}_k^\dagger\right] = \delta_{jk}. \qquad (73)$$

In this new basis, the energy matrix is diagonal, and (69) gives the energy matrix elements,

$$H_{jk} = (\tilde{\phi}_j | \tilde{\phi}_k) = \int d^3x \, \tilde{\phi}_j(\mathbf{x},t)^\dagger \tilde{\phi}_k(\mathbf{x},t) = \hbar \omega_j \delta_{jk}, \qquad (74)$$

where we have introduced the eigen-frequencies $\omega_j$, such that $\hbar \omega_j$ equals the energy eigenvalues. The energy eigenstates for each helicity obey the eigenvalue equation (restoring $\sigma$ in the notation)

$$\hat{H}_\sigma \tilde{\phi}_{j,\sigma}(\mathbf{x},t) = \hbar c \sigma \nabla \times \tilde{\phi}_{j,\sigma}(\mathbf{x},t) = \hbar \omega_j \tilde{\phi}_{j,\sigma}(\mathbf{x},t). \qquad (75)$$

One can easily verify that this eigenvalue equation (with the overlap given by (74)) is solved by the following mode functions

$$\tilde{\phi}_{j,\sigma}(\mathbf{x},t) = i\left(\frac{\hbar \omega_j}{2}\right)^{1/2} \left[\mathbf{u}_{j,\sigma}(\mathbf{x},t) + i\sigma \frac{\mathbf{k}_j}{k_j} \times \mathbf{u}_{j,\sigma}(\mathbf{x},t)\right], \qquad (76)$$

where the vector plane-wave functions $\mathbf{u}_j(\mathbf{x},t)$, are given by

$$\mathbf{u}_{j,\sigma}(\mathbf{x},t) = \frac{1}{\sqrt{V}} \mathbf{e}_{\mathbf{k}_j,\sigma} \exp\left[i\left(\mathbf{k}_j \cdot \mathbf{x} - \omega_j t\right)\right]. \tag{77}$$

Here the $\mathbf{k}_j$ are wave vectors orthogonal to the transverse circular-polarization unit vectors $\mathbf{e}_{\mathbf{k}_j,\sigma}$, and $V$ is the quantization volume. In the continuum limit, these functions go to those given in (4). The photon-field operator in (62) may thus be expressed as

$$\hat{\Psi}(\mathbf{x},t) = \sum_{j,\sigma} \tilde{\phi}_{j,\sigma}(\mathbf{x},t) \hat{a}_{j,\sigma} + H.c. \tag{78}$$

Note that the energy-eigenstate basis functions (76), are orthogonal not only under the non-local scalar product (59), but are also orthogonal with respect to the overlap integral, (74).[9] But they are not normalized to unity under (74). Therefore, we normalize them to unity by dividing by $(\hbar\omega_j)^{1/2}$

$$\phi_{j,\sigma}(\mathbf{x},t) = \tilde{\phi}_{j,\sigma}(\mathbf{x},t)(\hbar\omega_j)^{-1/2} = i\frac{1}{\sqrt{2}}\left[\mathbf{u}_{j\sigma}(\mathbf{x},t) + i\sigma\frac{\mathbf{k}_j}{k_j} \times \mathbf{u}_{j,\sigma}(\mathbf{x},t)\right]. \tag{79}$$

This leads to the following form for the photon-field operator

$$\hat{\Psi}(\mathbf{x},t) = \sum_{j,\sigma} (\hbar\omega_j)^{1/2} \phi_{j,\sigma}(\mathbf{x},t) \hat{a}_{j,\sigma} + H.c.$$

$$= i\sum_{j,\sigma} \left(\frac{\hbar\omega_j}{2}\right)^{1/2} \left[\mathbf{u}_{j,\sigma}(\mathbf{x},t) + i\sigma\frac{\mathbf{k}_j}{k_j} \times \mathbf{u}_{j,\sigma}(\mathbf{x},t)\right]\hat{a}_{j,\sigma} + H.c. \tag{80}$$

$$= i\sum_{j,\sigma} \left(\frac{\hbar\omega_j}{2V}\right)^{1/2} \left[\mathbf{e}_{\mathbf{k}_j,\sigma} + i\sigma\frac{\mathbf{k}_j}{k_j} \times \mathbf{e}_{\mathbf{k}_j,\sigma}\right] \exp\left[i\left(\mathbf{k}_j \cdot \mathbf{x} - \omega_j t\right)\right]\hat{a}_{j,\sigma} + H.c..$$

Separating the terms in this equation into "real" and "imaginary" parts, as in (68), shows that in the continuum limit this field operator is equivalent (within a factor $1/\sqrt{\varepsilon_0}$) to the monochromatic-mode expression of Dirac theory for the quantized electromagnetic field in (2) and (3). The positive-frequency electric and magnetic field operators are then given by

$$\hat{\mathbf{E}}^{(+)}(\mathbf{x},t) = i\sum_{j,\sigma} \left(\frac{\hbar\omega_j}{2\varepsilon_0 V}\right)^{1/2} \mathbf{e}_{\mathbf{k}_j,\sigma} \exp\left[i\left(\mathbf{k}_j \cdot \mathbf{x} - \omega_j t\right)\right]\hat{a}_{j,\sigma}, \tag{81}$$

and

$$\hat{\mathbf{B}}^{(+)}(\mathbf{x},t) = i\sum_{j,\sigma} \left(\frac{\hbar\omega_j}{2\varepsilon_0 V}\right)^{1/2} \left(\frac{\mathbf{k}_j}{ck_j} \times \mathbf{e}_{\mathbf{k}_j,\sigma}\right) \exp\left[i\left(\mathbf{k}_j \cdot \mathbf{x} - \omega_j t\right)\right]\hat{a}_{j,\sigma}. \tag{82}$$

We have thus shown that both the monochromatic Dirac and the polychromatic Titulaer-Glauber quantized field theories of electromagnetism can be derived from a photon-energy wave function in much the same way as one arrives at the quantum field theory for electrons. The photon wave function and its equation of motion are obtained by finding the first-order wave equation corresponding to the Einstein energy-momentum-mass relation for a massless, spin-1 particle. We derived the Poincaré-invariant scalar product for the photon wave function in coordinate space by requiring a local expression for the energy expectation value. The scalar product is non-local in the coordinate representation, but local in the momentum representation. We showed that this scalar product can be used to better interpret and utilize the Titulaer-Glauber wave-packet quantization formalism.

---

[9] Therefore, we conclude the energy eigenstates are proportional to their own dual modes, that is, from (19) they are eigenfunctions of the integral equation $\tilde{\phi}_{j,\sigma}(\mathbf{x},t) = \omega_j (2\pi^2 c\sigma)^{-1} \int \tilde{\phi}_{j,\sigma}(\mathbf{x}',t) |\mathbf{x}-\mathbf{x}'|^{-2} d^3 x'$. This is actually the integral representation of the differential eigenvalue equation $c\sigma \nabla \times \tilde{\phi}_{j\sigma}(\mathbf{x},t) = \omega_j \tilde{\phi}_{j\sigma}(\mathbf{x},t)$. This can be shown by direct calculation, and also follows from the fact that $|\mathbf{x}-\mathbf{x}'|^{-2}$ is the Green's function for the equation $c\sigma\sqrt{\nabla^2}\tilde{\phi}_{j\sigma}(\mathbf{x},t) = \omega_j \tilde{\phi}_{j\sigma}(\mathbf{x},t)$, (i.e. for transverse fields $\sqrt{\nabla^2} = \nabla \times$).

We have found, as have others [8-11, 43, 48, 49], that the photon wave function obeys a first-order equation of motion that is identical in form to the Maxwell equations (at least in free space). One should resist, however, interpreting this as evidence that the electric and magnetic fields for a single photon physically exist. We suggest that the photon wave function may be taken to be the fundamental object. In order for a quantity to "exist" in the sense we are using, it should be experimentally measurable given only one copy of a physical system. A wave function for a single object represents the quantum state of that object, and is not measurable, even in principle [50, 51]. This contrasts with the case where one has many replicas of the object, allowing quantum-state tomography to reconstruct the common state [52-54]. We suggest that the macroscopic electric and magnetic fields appear as emergent properties of a collection of many photons, in spirit similar to the discussion of Merzbacher, who does not invoke photon wave functions [17]. This is analogous to the emergence of a macroscopic spin associated with a collection of many atoms, whose collective state of spin can be determined by weak measurements on the entire ensemble [55-57]. In this sense, we say that the traditional electromagnetic field is described by a mean-field theory known as Maxwell's equations.

## 5. Modes versus states and the wave-function extraction rule

There is a one-to-one relationship between the modes used in quantum field theory (QFT) and the states used in photon wave mechanics (PWM). This relationship can be seen most easily in cases that we can restrict the field's state description to a fixed subspace of the larger Fock space of photon numbers. To demonstrate this relationship, we write out the states for a few examples commonly encountered in quantum optics. The simplest example is that of a single-photon pure state, which in QFT may be denoted by

$$\left|1_\phi\right\rangle = \hat{a}_\phi^\dagger \left|vac\right\rangle, \tag{83}$$

where $\hat{a}_\phi^\dagger$ is the creation operator for a single photon in the spatial-temporal wave-packet mode $\phi(\mathbf{x},t)$, analogous to (10) and (26). In photon wave mechanics, all information about this state is represented by the single-photon wave function,

$$\Psi^{(1)}(\mathbf{x},t) = \phi(\mathbf{x},t). \tag{84}$$

Equation (83) is the state of a field, while (84) is the state of a particle. Both are fully described by specifying the function $\phi(\mathbf{x},t)$, which plays the role of a mode function in QFT and a state function in PWM. In either case, the function $\phi(\mathbf{x},t)$ obeys the complex form of the Maxwell equations (45).

To see the equivalence of the two formalisms more generally, consider a single-photon pure state in terms of the plane-wave basis, in both QFT and PWM. In QFT, a single-photon state is given by

$$\left|\Psi^{(1)}\right\rangle = \sum_{\mathbf{k},\sigma} C_{\mathbf{k},\sigma} \hat{a}_{\mathbf{k},\sigma}^\dagger \left|vac\right\rangle, \tag{85}$$

where, as before, $\sigma$ labels the two circular-polarization states associated with a particular wave vector. The coefficients form one column of a unitary matrix, and so satisfy $\sum_{\mathbf{k},\sigma} C_{\mathbf{k},\sigma}^* C_{\mathbf{k},\sigma} = 1$. Calculation of the matrix element for the positive-frequency field operator $\hat{\Psi}_\sigma^{(+)}(\mathbf{x},t)$ between the vacuum and the single-photon state of the field gives

$$\Psi^{(1)}(\mathbf{x},t) = \langle \text{vac}| \hat{\Psi}_\sigma^{(+)}(\mathbf{x},t) |\Psi^{(1)}\rangle = \langle \text{vac}| \left(\frac{\varepsilon_0}{2}\right)^{1/2} \left(\hat{\mathbf{E}}^{(+)}(\mathbf{x},t) + i\sigma c \hat{\mathbf{B}}^{(+)}(\mathbf{x},t)\right)|\Psi^{(1)}\rangle$$

$$= \langle \text{vac}| \left(\frac{\varepsilon_0}{2}\right)^{1/2} i \sum_{\mathbf{k},\sigma} \left(\frac{\hbar\omega_\mathbf{k}}{2\varepsilon_0 V}\right)^{1/2} \left(\mathbf{e}_{\mathbf{k},\sigma} + i\sigma \frac{c\mathbf{k}}{ck} \times \mathbf{e}_{\mathbf{k},\sigma}\right) \exp[i(\mathbf{k}\cdot\mathbf{x} - \omega_\mathbf{k}t)] \hat{a}_{\mathbf{k},\sigma}|\Psi^{(1)}\rangle \quad (86)$$

$$= i \sum_{\mathbf{k},\sigma} C_{\mathbf{k},\sigma} \left(\frac{\hbar\omega_\mathbf{k}}{4V}\right)^{1/2} \left(\mathbf{e}_{\mathbf{k},\sigma} + i\sigma \frac{\mathbf{k}}{k} \times \mathbf{e}_{\mathbf{k},\sigma}\right) \exp[i(\mathbf{k}\cdot\mathbf{x} - \omega_\mathbf{k}t)].$$

This is of the form of the wave-packet modes in (10), and so equals a photon wave function (PWF).

Equation (86) shows how to extract a PWF from the QFT description of a one-photon state of the field [19], so we call it the "extraction rule." In fact, the rule is analogous to that used in QFT of electrons [34]. To illustrate this, rewrite (86) as

$$\langle \text{vac}| \hat{\Psi}_\sigma^{(+)}(\mathbf{x},t) |\Psi^{(1)}\rangle = \langle (\sigma,\mathbf{x})|\Psi^{(1)}(t)\rangle, \quad (87)$$

where we introduce the "photon reference states" by the definition

$$|(\sigma,\mathbf{x})\rangle = \hat{\Psi}_\sigma^{(-)}(\mathbf{x},0)|\text{vac}\rangle = \left(\frac{\varepsilon_0}{2}\right)^{1/2} \left(\hat{\mathbf{E}}^{(-)}(\mathbf{x},0) - i\sigma c \hat{\mathbf{B}}^{(-)}(\mathbf{x},0)\right)|\text{vac}\rangle$$

$$= -i\left(\frac{\varepsilon_0}{2}\right)^{1/2} \sum_{\mathbf{k},\lambda} \left(\frac{\hbar\omega_\mathbf{k}}{2\varepsilon_0 V}\right)^{1/2} \left(\mathbf{e}_{\mathbf{k},\lambda}^* - i\sigma \frac{\mathbf{k}}{k} \times \mathbf{e}_{\mathbf{k},\lambda}^*\right) \exp(-i\mathbf{k}\cdot\mathbf{x})|1\rangle_{\mathbf{k},\lambda}$$

$$= -i\left(\frac{\varepsilon_0}{2}\right)^{1/2} \sum_{\mathbf{k},\lambda} \left(\frac{\hbar\omega_\mathbf{k}}{2\varepsilon_0 V}\right)^{1/2} \left(\mathbf{e}_{\mathbf{k},\lambda}^* - i\sigma(i\lambda \mathbf{e}_{\mathbf{k},\lambda}^*)\right) \exp(-i\mathbf{k}\cdot\mathbf{x})|1\rangle_{\mathbf{k},\lambda} \quad (88)$$

$$= -i\sum_\mathbf{k} \left(\frac{\hbar\omega_\mathbf{k}}{V}\right)^{1/2} \mathbf{e}_{\mathbf{k},\sigma}^* \exp(-i\mathbf{k}\cdot\mathbf{x})|1\rangle_{\mathbf{k},\sigma}.$$

Here we have used the relationship (5), between the wave vector $\mathbf{k}$, and polarization vector $\mathbf{e}_{\mathbf{k},\sigma}$.

The states $|(\sigma,\mathbf{x})\rangle$ are not position eigenstates, but they play a role similar to that of such states for the case of electrons.[10] Their overlap integral is not a delta function, but is given by

$$\langle(\rho,\mathbf{x}')|(\sigma,\mathbf{x})\rangle = \frac{\hbar c \delta_{\rho\sigma}}{\pi^2 |\mathbf{x}-\mathbf{x}'|^4}. \quad (89)$$

This can be interpreted as the wave function for the reference state $|(\sigma,\mathbf{x})\rangle$, and is reasonably localized about $\mathbf{x}'=\mathbf{x}$, as we would like to have.

## 6. Two-photon and multiphoton wave mechanics

In this section, we generalize the PWM formalism to describe two-photon states (treated in [21, 58] and [19, 49]), as well as multiphoton states.

### 6.1. Two-photon wave mechanics

The photon wave function concept can be extended to more than one photon. In an earlier paper [21], we introduced a two-photon wave function more or less by postulate. In analogy with the two-electron case, we required that the wave function depend on six space variables and one time variable, and that it satisfy a wave equation whose time derivative is proportional to the sum of two Hamiltonian operators of the form (37). Here we verify that this postulate is supported by the PWF-extraction rule given in the previous section when generalized by including a product of two field operators evaluated at distinct points. An arbitrary two-photon state in QFT is given by the generalization of (85), that is,

$$\left|\Psi^{(2)}\right\rangle = \sum_{j,\sigma}\sum_{m,\rho} C_{(j,\sigma),(m,\rho)} \hat{b}^{\dagger}_{j,\sigma}\hat{b}^{\dagger}_{m,\rho}\left|\mathrm{vac}\right\rangle. \qquad (90)$$

---

[10] For electrons, the extraction rule is: $\langle \text{vac}|\hat{\Psi}_\sigma^{(+)}(\mathbf{x},0)|\Psi^{(1)}(t)\rangle = \langle \mathbf{x}|\Psi^{(1)}(t)\rangle = \psi(\mathbf{x},t)$. The reference states $|\mathbf{x}\rangle$ are position eigenstates in the non-relativistic limit.

The positive-frequency part of the field operator is, from (62),

$$\hat{\Psi}^{(+)}(\mathbf{x},t) = \sum_{j,\sigma} \hat{b}_{j,\sigma}\,\psi_{j,\sigma}(\mathbf{x},t). \qquad (91)$$

Then, the PWF-extraction rule for the two-photon state is:

$$\begin{aligned}\Psi^{(2)}(\mathbf{x}_1,\mathbf{x}_2,t) &= \left\langle\mathrm{vac}\right|\hat{\Psi}^{(+)}(\mathbf{x}_1,t)\hat{\Psi}^{(+)}(\mathbf{x}_2,t)\left|\Psi^{(2)}\right\rangle \\ &= \left\langle\mathrm{vac}\right|\sum_{r,\eta}\hat{b}_{r,\eta}\psi_{r,\eta}(\mathbf{x}_1,t)\sum_{s,\mu}\hat{b}_{s,\mu}\psi_{s,\mu}(\mathbf{x}_2,t)\sum_{j,\sigma}\sum_{m,\rho}C_{(j,\sigma),(m,\rho)}\hat{b}^{\dagger}_{j,\sigma}\hat{b}^{\dagger}_{m,\rho}\left|\mathrm{vac}\right\rangle \quad (92) \\ &= \sum_{r,\eta}\sum_{s,\mu}\left(C_{(r,\eta),(s,\mu)}+C_{(s,\mu),(r,\eta)}\right)\psi_{r,\eta}(\mathbf{x}_1,t)\psi_{s,\mu}(\mathbf{x}_2,t).\end{aligned}$$

We see that the correct bosonic symmetrization automatically emerges in the extracted PWF. The two-photon wave function is the energy-density amplitude for localizing the energies of the two photons at two different spatial points $\mathbf{x}_1$ and $\mathbf{x}_2$ at time $t$.

To simplify the notation, we will hereafter use the convention $j\equiv(j,\sigma)$, $m\equiv(m,\rho)$ to specify particular wave-packet modes or their creation operators. So, $C_{jm}$ means $C_{(j,\sigma),(m,\rho)}$. We also will incorporate the symmetrization implicitly into the sum, rather than display it explicitly as in (92). Then the two-photon wave function $\Psi^{(2)}(\mathbf{x}_1,\mathbf{x}_2,t)$ is written [19, 21, 58]

$$\Psi^{(2)}(\mathbf{x}_1,\mathbf{x}_2,t) = \sum_{j,m} C_{jm}\,\psi^{(1)}_j(\mathbf{x}_1,t)\otimes\psi^{(1)}_m(\mathbf{x}_2,t), \qquad (93)$$

where the coefficients $C_{jm}$ symmetrize the wave function under particle-label exchange, and $\otimes$ is the tensor product between the photon state spaces. Note that in keeping with standard QFT notation, the tensor product is suppressed in (92), but is nevertheless implicitly present. The modulus-squared of the expansion coefficients, $\left|C_{jm}\right|^2$, gives the probabilities of the photons being in the states labelled by $j$ and $m$. Each coefficient $C_{jm}$ can be determined by applying (61) twice to $\Psi^{(2)}(\mathbf{x}_1,\mathbf{x}_2,t)$. The basis states $\left\{\psi^{(1)}_m\right\}$ are solutions of single-photon wave equations, and they include the spin dependence through $m\equiv(m,\rho)$.

The vacuum equation of motion for the two-photon wave function, referred to as the two-photon wave equation [21, 58], is the sum of the Hamiltonians for the individual photons,

$$i\hbar\partial_t\Psi^{(2)} = \hbar c\,\alpha^{(2)}_1\nabla_1\times\Psi^{(2)} + \hbar c\,\alpha^{(2)}_2\nabla_2\times\Psi^{(2)}, \qquad (94)$$

where the differential operators are understood to act on the appropriate components of the tensor product, and

$$\alpha^{(2)}_1 = \Sigma_3\otimes\mathbf{I}, \quad \alpha^{(2)}_2 = \mathbf{I}\otimes\Sigma_3, \qquad (95)$$

where

$$\mathbf{I}=\begin{pmatrix}1 & 0 \\ 0 & 1\end{pmatrix},\,\Sigma_3=\begin{pmatrix}1 & 0 \\ 0 & -1\end{pmatrix},\,\mathbf{1}=\begin{pmatrix}1 & 0 & 0 \\ 0 & 1 & 0 \\ 0 & 0 & 1\end{pmatrix}. \qquad (96)$$

The two-photon wave function also obeys the zero-divergence condition

$$\nabla_j\cdot\Psi^{(2)} = 0, \qquad (j=1,2), \qquad (97)$$

in which the differential operator acts on the appropriate tensor component. We call (94) the *two-photon wave equation*.

Tracing over the tensor product of the Hermitian conjugate of the two-photon wave function with itself, and integrating over all space, gives the expectation value of the product of the two photons' energies

$$\iint Tr\left[\Psi^{(2)\dagger}\Psi^{(2)}\right]d^3x_1 d^3x_2 = \langle E_1 E_2 \rangle. \tag{98}$$

If the state of the two photons is not entangled, this simplifies to $\langle E_1 E_2 \rangle = \langle E_1 \rangle \langle E_2 \rangle$.

The influence of an inhomogeneous linear refractive index on the propagation of photon wave functions can be treated phenomenologically by modifying the wave equation, as described in [10, 21, 58].

We propose that, just as the physical electromagnetic fields **E** and **B** emerge only in the limit that many statistically independent photons are present, a physical biphoton field emerges if many statistically independent photon pairs are present. Such a situation occurs as a result of parametric down conversion, in which individual blue photons can be spontaneously converted into pairs of red photons [22].

*6.2. Two-time wave functions*

There is an alternative way to define a two-photon wave function by using two distinct times. This is useful in the discussions of coherence theory in the following section. Early discussions of two-time wave functions can be found in [59-62]. In (92), instead of assuming a common time in the field operators, use two distinct times, $t_1, t_2$, as is sometimes done in quantum optics [22, 23, 25]. This generates the two-time, two-photon wave function,

$$\Phi^{(2)}(\mathbf{x}_1,t_1;\mathbf{x}_2,t_2) = \sum_{j,m} C_{jm} \psi_j^{(1)}(\mathbf{x}_1,t_1) \otimes \psi_m^{(1)}(\mathbf{x}_2,t_2). \tag{99}$$

As a consequence of the single-photon, one-time equation of motion (45), this two-time wave function simultaneously obeys a pair of wave equations

$$i\hbar\frac{\partial}{\partial t_j}\Phi^{(2)}(\mathbf{x}_1,t_1;\mathbf{x}_2,t_2) = \hbar c \alpha_j^{(2)} \nabla_j \times \Phi^{(2)}(\mathbf{x}_1,t_1;\mathbf{x}_2,t_2), \quad (j=1,2), \tag{100}$$

and the zero-divergence conditions

$$\nabla_j \cdot \Phi^{(2)}(\mathbf{x}_1,t_1;\mathbf{x}_2,t_2) = 0, \quad (j=1,2). \tag{101}$$

The two-time, two-photon formalism may also be arrived at in a manner similar to that for the one-time formalism through a pair of energy-momentum-mass equations analogous to (27).

Here we show that these two descriptions [(93) and (99)] of the two-photon state are equivalent under the standard measurement-collapse hypothesis of quantum mechanics. Consider a two-photon state described by the one-time wave function in (93). Suppose that a point-like photo-detector signals a detection event at time $T_1$ and position $\mathbf{R}_1$. After gaining this new knowledge, an observer should collapse his or her state description, (93), to the state

$$\Psi^{(2)}(\mathbf{x}_1,\mathbf{x}_2,t) \rightarrow \Psi^{(2)}(\mathbf{R}_1,T_1;\mathbf{x},t) = \sum_{j,m} C_{jm} \psi_j^{(1)}(\mathbf{R}_1,T_1) \otimes \psi_m^{(1)}(\mathbf{x},t), \tag{102}$$

If we interpret $T_1$ as $t_1$ and $t$ as $t_2$, we see that (102) has the same form as (99). Therefore, they obey the same wave equation, (100). We see that the two-time wave equation evolves not in the "absolute wall-clock" time variable $t$, but rather in the "measurement times" $t_1, t_2$. This illustrates that two-time wave functions can be used for predicting correlations between measurements at distinct space-time points. It also shows that state collapse is simply one, non-essential, method for describing these correlations. These conclusions hold for states of electrons as well as photons.

*6.3. Two-photon mixed states*

In the case of pure states, the two-photon wave function contains all obtainable knowledge about the state of the two-photon system. For the case of non-pure or mixed states, one can construct two-photon density matrices from two-photon wave functions as one does in standard quantum mechanics [63]. However, when one wishes to calculate the reduced density matrix for a system of photons, by tracing out all information about one photon, confusion may arise when choosing the form of the tracing operation [19, 64]. As an example, consider a two-photon pure state described by (93). The density matrix for this state may be written as

$$\rho^{(2)}\left(\mathbf{x}_1,\mathbf{x}_1',\mathbf{x}_2,\mathbf{x}_2',t\right) = \Psi^{(2)}(\mathbf{x}_1,\mathbf{x}_2,t)\Psi^{(2)}\left(\mathbf{x}_1',\mathbf{x}_2',t\right)^\dagger. \quad (103)$$

Suppose that we discard all information about the photon labelled by position 2. In standard quantum mechanics, we would describe the remaining photon state by the reduced density matrix, formed by setting $\mathbf{x}_2' = \mathbf{x}_2$, and integrating over $\mathbf{x}_2$. However, we observe that this is not the proper way to eliminate all information about photon 2 [64], because this uses the wrong scalar product, and does not eliminate the units of energy related to photon 2. In fact, such a procedure retains information about photon 2. To extinguish all information regarding photon 2, one must use the proper scalar product, which for the case of photons, is the non-local expression in coordinate space (53). Thus the proper reduced density matrix for photon 1 is given by

$$\rho^{(1)}\left(\mathbf{x}_1,\mathbf{x}_1',t\right) = \frac{1}{2\pi^2\hbar c}\int d^3x_2 \int d^3x_2' \frac{\Psi^{(2)}(\mathbf{x}_1,\mathbf{x}_2,t)\Psi^{(2)}\left(\mathbf{x}_1',\mathbf{x}_2',t\right)^\dagger}{\left|\mathbf{x}_2-\mathbf{x}_2'\right|^2}. \quad (104)$$

This density-matrix description and trace operation is readily extended to multiphoton states.

*6.4. Multiphoton wave mechanics*

The two-photon wave mechanics developed above can be extended to any number of photons. A $n$-photon pure state may be characterized by a $n$-photon, one-time wave function given by

$$\Psi^{(n)}(\mathbf{x}_1,\mathbf{x}_2,\cdots,\mathbf{x}_n;t) = \sum_{\{a\}} C_{\{a\}} \bigotimes_{j=1}^{n} \psi_{\{a\}_j}^{(1)}(\mathbf{x}_j,t), \quad (105)$$

where the $\left\{\psi_l^{(1)}\right\}$ are a set of single-photon basis states, the $C_{\{a\}}$ are expansion coefficients that symmetrize the state with respect to particle-label exchange, the sum is taken over the set of basis element labels $\{a\} = \{a_1,a_2,\cdots,a_n\}$, and $\{a\}_j$ is the $j$th entry of the basis element label $\{a\}$. The tensor product is taken over the $n$ different single-photon states, evaluated at different space coordinates. The equation of motion for this $n$-photon wave function is found by adding the Hamiltonians associated with the different photon coordinates,

$$i\hbar\partial_t \Psi^{(n)}(\mathbf{x}_1,\mathbf{x}_2,\cdots,\mathbf{x}_n;t) = \sum_{j=1}^{n} \hbar c \alpha_j^{(n)} \nabla_j \times \Psi^{(n)}(\mathbf{x}_1,\mathbf{x}_2,\cdots,\mathbf{x}_n;t), \quad (106)$$

where the $\alpha_j^{(n)}$ are the straightforward generalization of the matrices defined in (95). As in the two-photon case, there exists a generalized $n$-time, $n$-photon wave function, which is related to the one-time, $n$-photon wave function defined above, through the standard measurement-collapse hypothesis of quantum mechanics as described in Section 6.2. It is straightforward to see this relationship, and we will not go though the argument. The reduced density matrices for subsets of photons are obtained in the same manner as for the two-photon case, with a double, non-local integration for each photon to be eliminated.

Recall that there is a direct correspondence between the modes of QFT and the states of single-photon wave mechanics, which we discussed in Section 5. To extend this to a multi-mode $n$-photon pure state, write the state of the field as

$$\left|\Psi^{(n)}\right\rangle = \sum_{\{b\}} C_{\{b\}} \bigotimes_{j=1}^{n} \left|\psi_{\{b\}_j}\right\rangle. \quad (107)$$

In the PWM description, the state of the photons is

$$\Psi^{(n)}(\mathbf{x}_1,\mathbf{x}_2,\cdots,\mathbf{x}_n;t) = \sum_{\{b\}} C_{\{b\}} \bigotimes_{j=1}^{n} \psi_{\{b\}_j}^{(1)}(\mathbf{x}_j,t). \quad (108)$$

It is interesting to note that these two equations have the same structure. Indeed this is due to the fact that they are in effect, equivalent theories. In the Schrödinger picture of quantum mechanics, the states given by (107) evolve in the same way as does the $n$-photon wave function in (108). An alternative treatment of many-photon wave mechanics is given in [20].

## 7. Relation of photon wave mechanics to the Wolf equations of classical coherence theory

A strong argument in favour of the energy-density wave function form of PWM is that it bears strong connections to other, well-established theories—both quantum and classical—such as photo-detection theory, classical and quantum optical coherence theory [22, 44, 65-67], and the biphoton amplitude, which is used in most discussions of spontaneous parametric down conversion [68-71]. In this section we develop a close connection between two-photon wave mechanics, described by the two-photon wave functions (both one-time and two-time), and second-order optical coherence theory. We first write the second-order correlation functions and their equations of motion. The most general second-order coherence description of the electromagnetic field in the context of semi-classical theory, as defined by Mandel and Wolf, is given by the second-order coherence matrices (tensors) [22, 72]

$$\begin{aligned}
\mathbf{E}(x_1, x_2) &= \left\langle \mathbf{E}^{(-)}(x_1) \otimes \mathbf{E}^{(+)}(x_2) \right\rangle = \left[ \left\langle E_r^{(-)}(x_1) E_s^{(+)}(x_2) \right\rangle \right], \\
\mathbf{H}(x_1, x_2) &= \left\langle \mathbf{H}^{(-)}(x_1) \otimes \mathbf{H}^{(+)}(x_2) \right\rangle = \left[ \left\langle H_r^{(-)}(x_1) H_s^{(+)}(x_2) \right\rangle \right], \\
\mathbf{M}(x_1, x_2) &= \left\langle \mathbf{E}^{(-)}(x_1) \otimes \mathbf{H}^{(+)}(x_2) \right\rangle = \left[ \left\langle E_r^{(-)}(x_1) H_s^{(+)}(x_2) \right\rangle \right], \\
\mathbf{N}(x_1, x_2) &= \left\langle \mathbf{H}^{(-)}(x_1) \otimes \mathbf{E}^{(+)}(x_2) \right\rangle = \left[ \left\langle H_r^{(-)}(x_1) E_s^{(+)}(x_2) \right\rangle \right],
\end{aligned} \quad (109)$$

where $\mathbf{E}^{(\pm)}, \mathbf{H}^{(\pm)}$ are the positive- and negative-frequency components of the electric and magnetic field vectors, $x_j = (\mathbf{x}_j, t_j)$ are space-time coordinates, and $r, s \in \{x, y, z\}$ label the Cartesian components of the electric- and magnetic-field vectors. The notation $[\;]$ indicates a 3×3 matrix of correlation functions. These four matrices completely describe the correlations between various components of the electric and magnetic fields at two different space-time coordinates. In particular, they give a complete description of second-order partial coherence of an optical field (including spatial, temporal and polarization coherence). The magnetic field $\mathbf{H}^{(\pm)}$ is used rather than the magnetic induction $\mathbf{B}^{(\pm)}$ in order to accommodate linear-response materials into the treatment (although we will not develop that here). See BJS [58] and IBB [10].

We find from the Maxwell equations that the evolution of the coherence matrices is governed by a set of first-order linear differential equations, which we name the "first-order-in-time Wolf equations,"[11]

$$\nabla_j \times \mathbf{A} + \frac{1}{c} \frac{\partial}{\partial t_j} \mathbf{B} = \mathbf{0}, \quad \nabla_j \cdot \mathbf{F} = \mathbf{0}, \quad (j = 1, 2), \quad (110)$$

where $(\mathbf{A}, \mathbf{B})$ denote the matrix pairs $(\mathbf{A}, \mathbf{B}) \in \{(\mathbf{E}, -\mathbf{N}), (\mathbf{M}, -\mathbf{H}), (\mathbf{N}, \mathbf{E}), (\mathbf{H}, \mathbf{M})\}$, and the matrix $\mathbf{F}$ is any of the four coherence matrices. Here the curl and divergence are understood to act on the appropriate vector in the tensor-product. Equations (110) completely describe the evolution of the second-order coherence of an optical field as it propagates through free space. Consequences of an inhomogeneous medium can be incorporated without trouble [21, 58].

Using these equations, one can show that each component of the coherence matrices obeys the Wolf equations [22, 73], a well-known set of second-order-in-time differential equations

$$\left( \nabla_j^2 - \frac{1}{c^2} \frac{\partial^2}{\partial t_j^2} \right) F_{rs}(x_1, x_2) = 0, \quad (j = 1, 2), \quad (111)$$

---

[11] These equations generalize those in [22, 72, 73], which describe a time-stationary field.

where $\mathbf{F}_{rs}$ is the $r,s$ component of any of the second-order coherence matrices. The Wolf equations were recently highlighted for their relation to the two-photon detection amplitude [74], discussed below. Our treatment in the following generalizes that relation.

We combine the second-order coherence matrices into a single complex matrix, which is identical in form to the two-photon wave function,

$$\Gamma^{(2)}(x_1,x_2) = \begin{pmatrix} \gamma^{(2)}_{+1+1}(x_1,x_2) & \gamma^{(2)}_{+1-1}(x_1,x_2) \\ \gamma^{(2)}_{-1+1}(x_1,x_2) & \gamma^{(2)}_{-1-1}(x_1,x_2) \end{pmatrix}, \quad (112)$$

where the block matrix elements $\gamma^{(2)}_{\sigma_1\sigma_2}$, labelled by the helicity (circular polarization) $\sigma_1, \sigma_2$ of the fields at space-time points $x_1$ and $x_2$, are given by

$$\gamma^{(2)}_{\sigma_1\sigma_2}(x_1,x_2) = \frac{\varepsilon_0}{2}\left\{\mathbf{E}(x_1,x_2) + \frac{i}{c}\left[\sigma_1\mathbf{N}(x_1,x_2) + \sigma_2\mathbf{M}(x_1,x_2)\right] - \mu_0\sigma_1\sigma_2\mathbf{H}(x_1,x_2)\right\}. \quad (113)$$

We call this new complex matrix $\Gamma^{(2)}(x_1,x_2)$, the *second-order Riemann-Silberstein coherence matrix*, since it can be easily derived from the complex RS vector, (68). Each block matrix element may be written as an expectation value of the tensor product of two different RS vectors

$$\gamma^{(2)}_{\sigma_1\sigma_2}(x_1,x_2) = \left\langle \mathbf{F}^{(-)}_{-\sigma_1}(x_1) \otimes \mathbf{F}^{(+)}_{\sigma_2}(x_2) \right\rangle, \quad (114)$$

where

$$\mathbf{F}^{(+)}_\sigma(x_1) = \left(\frac{\varepsilon_0}{2}\right)^{1/2}\left[\mathbf{E}^{(+)}(x_1) + ic\sigma\mathbf{B}^{(+)}(x_1)\right]. \quad (115)$$

The negative sign of the helicity, $-\sigma_1$, on the first RS vector in (114) ensures that the RS coherence tensor (112) evolves in the same way as does the two-photon wave function. This relationship between two-photon wave mechanics and second-order optical coherence theory shows the equivalence of the evolution of two-photon states and second-order optical coherence. For a two-photon field, all higher-order coherence functions equal zero, which emphasizes why the second-order coherence functions contain all information about the state of the light.

In vacuum, the equations of motion for the second-order RS coherence matrix follow from the Maxwell equations, and are identical to the evolution equations describing the two-time, two-photon wave function, (100) and (101), that is,

$$i\hbar\frac{\partial}{\partial t_j}\Gamma^{(2)}(x_1,x_2) = \hbar c\alpha^{(2)}_j \nabla_j \times \Gamma^{(2)}(x_1,x_2), \quad (j=1,2), \quad (116)$$

and

$$\nabla_j \cdot \Gamma^{(2)}(x_1,x_2) = 0, \quad (j=1,2). \quad (117)$$

In a linear medium, these equations are modified in the same way as those for the two-photon wave function as described in [10, 21, 58]. From this result we see that the evolution of two-photon states is identical to that of the second-order optical correlation functions. Thus one can make use of the well-developed theories of second-order optical coherence propagation to describe the behaviour of two-photon states, as done in [21], where decoherence of a pair of spatially-entangled photons was modelled for propagation through a realistic atmosphere.

The two-time wave-function description has another relationship with classical coherence theory. Starting from the two-time equations of motion, (100), divide by $i\hbar c$ and take the time derivative of both sides, to give

$$\frac{1}{c^2}\frac{\partial^2}{\partial t_j^2}\Phi^{(2)}(\mathbf{x}_1,t_1;\mathbf{x}_2,t_2) = -i\nabla_j \times \left(\frac{1}{c}\frac{\partial}{\partial t_j}\Phi^{(2)}(\mathbf{x}_1,t_1;\mathbf{x}_2,t_2)\right) \quad (j=1,2). \quad (118)$$
$$= -\nabla_j \times \nabla_j \times \Phi^{(2)}(\mathbf{x}_1,t_1;\mathbf{x}_2,t_2) = \nabla_j^2 \Phi^{(2)}(\mathbf{x}_1,t_1;\mathbf{x}_2,t_2),$$

We have made use of the vector identity $\nabla \times \nabla \times \mathbf{F} = -\nabla^2\mathbf{F} + \nabla(\nabla\cdot\mathbf{F})$ and the fact that the wave function has zero divergence. Equations (118) are precisely the Wolf equations (111), which have been previously discussed in relation to two-photon detection amplitudes [74]. Note that the Wolf equations can be derived from the two-photon wave equations, but the converse is not true. This is analogous to the relationship between the Dirac equation and the Klein-Gordon equation for electrons. The two-photon wave equations contain more information about the evolution of the two-photon field than do the Wolf equations.

## 8. Relation of PWM to two-photon detection amplitudes

The two-photon detection amplitude $A_D^{(2)}(x_1, x_2)$ of standard quantum optics is proportional to the joint probability amplitude for detecting one photon in a volume much larger than a cubic wavelength, centred at the space-time coordinate $x_1$, and the second photon in a volume centred at $x_2$. Assuming the use of conventional photodetectors, which respond only to the electric field, the two-photon detection amplitude is given by [22, 25, 44, 65-67]

$$A_D^{(2)}(x_1, x_2) = \langle vac | \hat{\mathbf{E}}^{(+)}(x_1) \otimes \hat{\mathbf{E}}^{(+)}(x_2) | \Psi^{(2)} \rangle. \tag{119}$$

Here $\hat{\mathbf{E}}^{(+)}(x_j)$ is the positive-frequency part of the electric-field operator evaluated at the space-time coordinate $x_j$ of detector $D_j$, $(j = 1, 2)$. This has also been called the biphoton amplitude in the case of entangled photon pairs created by spontaneous parametric down conversion [68-71].

The two-photon detection amplitude is proportional to the real part of the two-photon wave function defined in (99) above. To see this more explicitly, note that the single-photon wave functions, from which the two-photon wave function is constructed, can be linked to the classical Maxwell equations through (65). There the real part of the single-photon wave function was linked with the electric field and the imaginary part with the magnetic field. Following this identification in the two-photon case, we see that the two-photon wave function may be written as

$$\Psi^{(2)}_{\sigma_1 \sigma_2}(\mathbf{x}_1, \mathbf{x}_2, t) = \frac{\varepsilon_0}{2} \left[ \mathbf{E}_1^{(+)} \mathbf{E}_2^{(+)} - \sigma_1 \sigma_2 c^2 \mathbf{B}_1^{(+)} \mathbf{B}_2^{(+)} + ic \left( \sigma_2 \mathbf{E}_1^{(+)} \mathbf{B}_2^{(+)} + \sigma_1 \mathbf{B}_1^{(+)} \mathbf{E}_2^{(+)} \right) \right]. \tag{120}$$

Here $\mathbf{E}_j^{(+)} \left( \mathbf{B}_j^{(+)} \right)$ is formally identified with the positive-frequency part of the (non-operator) electric (magnetic-induction) field, evaluated at the space-time coordinate $x_j$, and a tensor product is implicit in each **E** and **B** field product. Thus, if our detector is insensitive to magnetic fields, the real part of the two-photon wave function is identical in form to the two-photon detection amplitude. The same point has been made in [18]. This is not surprising since the photon-wave functions are based upon energy localization, and it is the energy that activates photon-counting detectors. The energy from photons is absorbed by such detectors and transformed into a photo-current of electrons. It should be clear that the electric-field operators in (119) evolve according to the Maxwell equations, as do the components of the two-photon wave function (120). Indeed, one can carry this full circle and note that one may define a two-photon detection amplitude based on the full electromagnetic field, which is identical in form to the two-photon wave function [21]. This idea would prove useful if the detector absorbs light by exciting a magnetic dipole.

## 9. Examples and applications

### 9.1. Example wave-packet modes

In this section, we develop a simple example of the non-monochromatic wave-packet modes introduced in Sections 2 and 3 to describe localized electromagnetic fields and localized single-photon states. As we pointed out in Section 5, wave-packet modes for the field are equivalent to states for the photon. In typical laboratory experiments, beam-like radiation is commonly encountered, that is, light propagating in primarily a single direction, and spatially confined in the plane transverse (perpendicular) to the propagation direction. In addition, this beam may consist of a train of pulses moving along the beam axis. The usual treatment of this geometry makes use of the paraxial and slowly-varying-temporal-envelope approximations [75].

Here we construct a set of spatial-temporal wave packets, which are approximately orthonormal under the Lorentz invariant scalar product (25) and (59). To do so we choose a form of the unitary transformation matrices $U_j^{(\sigma)}(\mathbf{k})$ in (10) to be given by

$$U_j^{(\sigma)}(\mathbf{k}) = \tilde{\psi}_l(k_x)\tilde{\psi}_m(k_y)\tilde{\psi}_n(k_z - \bar{k}_z), \tag{121}$$

where $j$ labels the triple indices $l, m, n$, and $\tilde{\psi}_m(k_i)$ is a normalized Hermite-Gaussian function,

$$\tilde{\psi}_m(k_i) = \left(\frac{\pi^{1/2} w_i}{2^{m-1} m!}\right)^{1/2} H_m(w_i k_i) \exp(-w_i^2 k_i^2 / 2), \qquad (i = x, y, z). \tag{122}$$

Here $1/w_i$ is the width (or spread) of the wave packet in momentum space in the $i$ th direction. For beam-like modes traveling in the $z$ direction, the longitudinal component of $\mathbf{k}$ denoted by $\bar{k}_z$, is much greater than $1/w_x$, and $1/w_y$. The coordinate-space wave-packet modes and their dual modes are given by the weighted Fourier transforms of the form in (10) and (19)

$$\mathbf{v}_{j,\sigma}(\mathbf{x},t) = i\left(\frac{\hbar c}{2\varepsilon_0}\right)^{1/2} \int \frac{d^3k}{(2\pi)^3} \sqrt{k}\, U_j^{(\sigma)}(\mathbf{k}) \mathbf{e}_{\mathbf{k},\sigma} \exp\left[i(\mathbf{k}\cdot\mathbf{x} - \omega_{\mathbf{k}} t)\right], \tag{123}$$

and

$$\mathbf{v}_{j,\sigma}^D(\mathbf{x},t) = i\left(\frac{2\varepsilon_0}{\hbar c}\right)^{1/2} \int \frac{d^3k}{(2\pi)^3} \frac{1}{\sqrt{k}} U_j^{(\sigma)}(\mathbf{k}) \mathbf{e}_{\mathbf{k},\sigma} \exp\left[i(\mathbf{k}\cdot\mathbf{x} - \omega_{\mathbf{k}} t)\right]. \tag{124}$$

In the paraxial limit the polarization vector $\mathbf{e}_{\mathbf{k},\sigma}$ does not vary with $\mathbf{k}$, therefore we factor it out of the integrals. So hereafter we write the mode functions as scalar functions. Similarly, in the paraxial approximation, the magnitude of the wave vector is dominated by the longitudinal $(z)$ component, and we can thus approximate the weight function $\sqrt{k} \approx |k_z|^{1/2}$. With this paraxial approximation, the transverse integrals factor and lead to the standard Hermite-Gaussian transverse spatial modes [76]. The longitudinal wave-packet modes, equivalent to the temporal wave packets, are given by an integral of the form

$$\psi_m(t_R) = i \int \frac{dk_z}{2\pi} |k_z|^{1/2} \tilde{\psi}_m(k_z - \bar{k}_z) \exp(-ik_z c t_R), \tag{125}$$

where $t_R = t - z/c$ is the retarded time. The corresponding longitudinal dual modes are given by a similar integral with $\sqrt{k}$ replaced by $|k_z|^{-1/2}$, (we absorb the constant prefactors $(\hbar c/2\varepsilon_0)^{1/2}$ and $(2\varepsilon_0/\hbar c)^{1/2}$ into the transverse mode functions). In this paraxial approximation, these integrals can be evaluated analytically in terms of hypergeometric functions. Figure 1 illustrates the real and imaginary parts of the longitudinal mode function and the corresponding dual-mode function for the mode index $m = 2$. Note that the mode functions have characteristic fast oscillations and a slowly-varying envelope.

Figure 2 shows the real and imaginary parts of the product of the complex conjugate of a dual mode with another wave packet mode for the same mode index $(m = 2)$ and for different mode indices $(m = 2$ and $m = 1)$. Notice that only the real part of the product of the dual-mode conjugate and its corresponding mode function gives a positive-definite result. The fact that the imaginary parts are non-zero indicates that even though these give the proper normalization, there is not a local photon-number density. The orthonormality of these wave-packet modes and their corresponding dual modes were checked by numerical integration and do indeed converge correctly. The "instantaneous" overlap is non-zero in either case, but when integrated over all time only the real part of the overlap between the mode function and its corresponding dual mode gives a non-zero result.

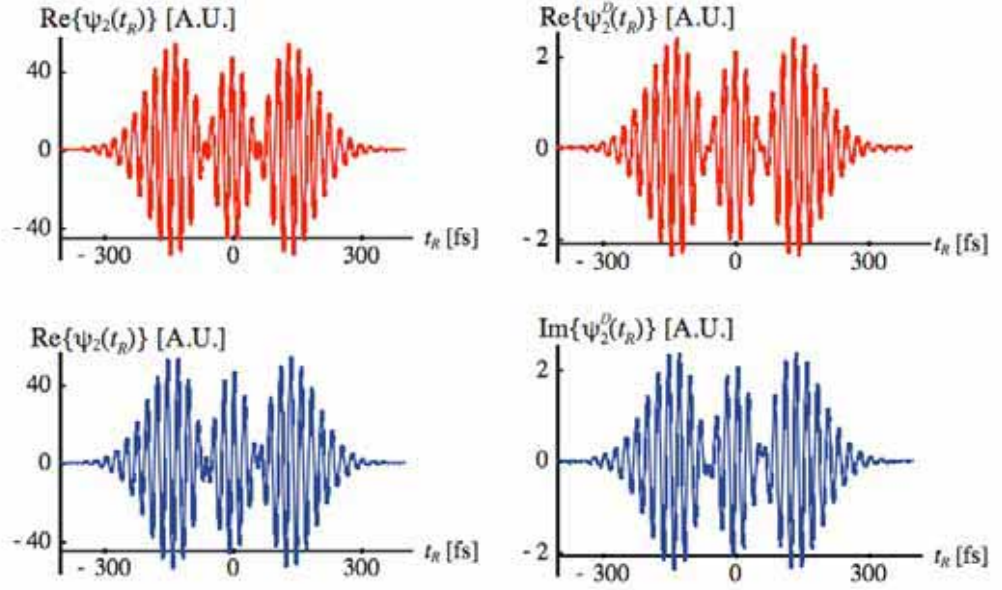

**Figure 1.** Real (top) and imaginary (bottom) parts of the $m=2$ "Hermite-Gaussian-like" wave-packet mode (left) and its corresponding dual mode (right) plotted as a function of the retarded time $t_R$. In each plot a central wavelength $\bar{\lambda} = 2\pi/\bar{k}_z = 810$ nm and temporal width $\tau = w_z/c = 60$ fs were used.

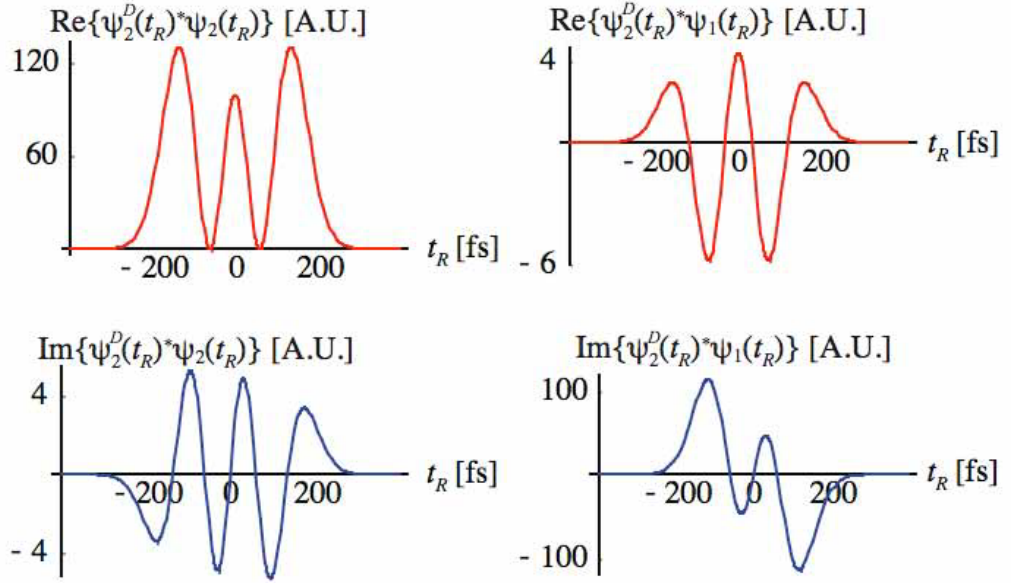

**Figure 2.** Real (top) and imaginary (bottom) parts of the product of the $m=2$ dual mode and the $m=2$ (left), $(m=1)$ (right), wave-packet mode $\psi_2^D(t_R)^* \psi_2(t_R)$, $\left(\psi_2^D(t_R)^* \psi_1(t_R)\right)$, as a function of retarded time $t_R$. In each plot a central wavelength $\bar{\lambda} = 2\pi/\bar{k}_z = 810$ nm and temporal width $\tau = w_z/c = 60$ fs were used.

Note that in this paraxial approximation, there is no coordinate-space relationship between the dual-mode functions and the mode functions analogous to (22). This can be viewed as a consequence of the one-dimensional nature of (125). Another result of the lack of a coordinate-space relation akin to (22) is that in one dimension there is, strictly speaking, no well-defined non-local scalar product in coordinate space. However, as we have just shown, the dual modes are well defined and their overlap with the corresponding wave-packet modes leads to a clear description of pulsed light. As pointed out in section 5, these mode functions can be used to expand the electromagnetic-field-operators in the quantum-field-theory description of light or to expand the single-photon wave functions in photon wave mechanics.

*9.2. Conversion between modes and dual modes*

As a second example, we consider how to convert physically from the wave-packet-mode basis to the corresponding dual-mode basis. For this one must have an apparatus that performs the non-local transformation in (22) or the corresponding transformation in the momentum representation. For beam-like modes, this amounts to spectral filtering. To determine the momentum-space transform needed, note that the Fourier amplitudes for the wave-packet and related dual modes are, from (123) and (124),

$$\tilde{\mathbf{v}}_{j,\sigma}(\mathbf{k},t) = i\left(\frac{\hbar ck}{2\varepsilon_0}\right)^{1/2} U_j^{(\sigma)}(\mathbf{k}) \mathbf{e}_{\mathbf{k},\sigma} \exp(-i\omega_\mathbf{k} t), \tag{126}$$

and

$$\tilde{\mathbf{v}}_{j,\sigma}^D(\mathbf{k},t) = i\left(\frac{2\varepsilon_0}{\hbar ck}\right)^{1/2} U_j^{(\sigma)}(\mathbf{k}) \mathbf{e}_{\mathbf{k},\sigma} \exp(-i\omega_\mathbf{k} t). \tag{127}$$

From these relations we see that, in momentum space, a dual mode is simply related to its corresponding wave-packet mode through multiplication by $2\varepsilon_0/\hbar ck$, that is,

$$\tilde{\mathbf{v}}_{j,\sigma}^D(\mathbf{k},t) = \frac{2\varepsilon_0}{\hbar c}\frac{\tilde{\mathbf{v}}_{j,\sigma}(\mathbf{k},t)}{k}. \tag{128}$$

Note that the influence of the $1/k$ factor is significant only for ultra-wide-band modes, that is, ultrashort wave packets.

From this expression it is straightforward to implement a dual-mode converter, which converts wave-packet modes into the corresponding dual modes, with non-unity efficiency. For beam-like paraxial modes one needs a spectral filter with transmission proportional to $1/\omega$. For sub-picosecond pulses this can be implemented by the optical system depicted in Figure 3. This system is a specialized version of the standard pulse shaper [75, 77]. It consists of two gratings (G1 and G2), two lenses (L1 and L2) with focal length $f$, and an amplitude mask (AM). The input wave-packet mode $\mathbf{v}_{j,\sigma}(\mathbf{x},t)$ is Fourier transformed by the grating and lens combination, so that in the Fourier transform (FT) plane of the lens, the wave-packet mode as a function of transverse position $(x)$, is proportional to the Fourier transform of the WP mode given by $\mathbf{v}_{j,\sigma}^{FT-plane}(x) \propto \tilde{\mathbf{v}}_{j,\sigma}(\omega x/(2\pi cf))$. Here $x$ is the transverse position, which in the Fourier plane of lens L1 corresponds to the frequency of the mode, and $\tilde{\mathbf{v}}_{j,\sigma}$ is the Fourier transform of the wave-packet mode. If we insert an amplitude modulator that multiplies by a factor $1/x$, we obtain the dual mode at the output, that is,

$$\mathbf{v}_{out}(\mathbf{x},t) \propto \mathbf{v}_{j,\sigma}^D(\mathbf{x},t). \tag{129}$$

The transverse position $x$ is defined to be positive definite, with $x=0$ defined as an offset corresponding to zero frequency (or zero energy for the photon).

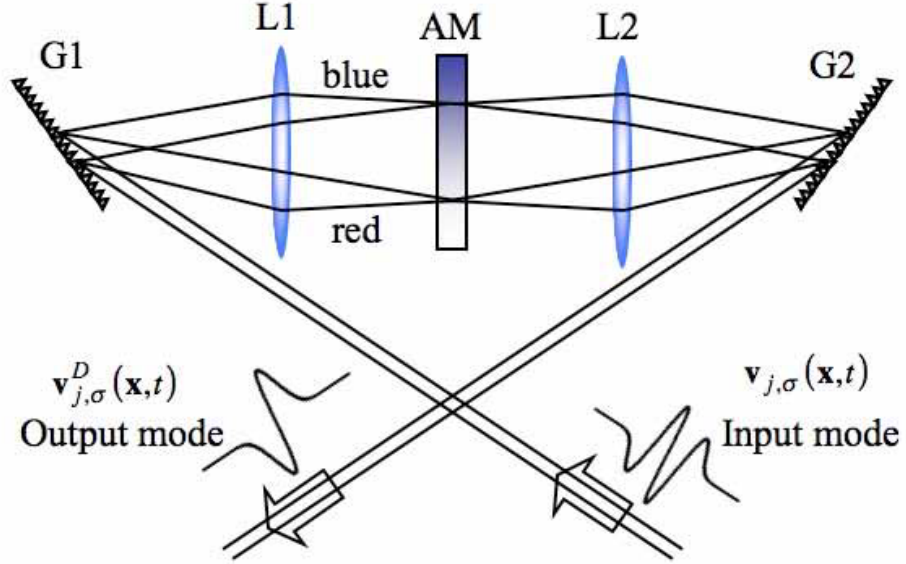

**Figure 3.** Dual-mode converter based upon a standard pulse shaper. Conversion occurs via spectral filtering in a dispersion-free grating-lens set up.

One may thus use such a universal mode converter to create a wave-packet dual mode from any WP mode, without needing to reprogram the shaper. Then one can interfere the dual mode with its original wave-packet on a standard square-law detector to measure the non-local scalar product. A precise procedure for such interference detection is balanced-homodyne detection, reviewed in [53]. The interference term in such an experiment is proportional to the non-standard scalar product defined in (25).

*9.3. Decoherence of spatially-entangled photon pair by atmospheric turbulence*

As a third example we consider the propagation through a turbulent atmosphere of two quasi-monochromatic photons, initially entangled in their transverse spatial degrees of freedom. In [21, 58] we assumed the photons are emitted in opposite directions occupying one of two orbital-angular-momentum (OAM) states, described by the standard, Laguerre-Gauss wave functions in the paraxial approximation. We assumed that the photons, labeled $A$ and $B$, have the same polarization and radial quantum number, and considered orbital quantum numbers of equal magnitudes separately. We treated the entangled input state $\Psi^{(2)} = \psi^A_{p,l} \otimes \psi^B_{p,-l} + \psi^A_{p,-l} \otimes \psi^B_{p,l}$, where $l$ is the OAM quantum number, and $p$ is the radial quantum number.

The photons pass through independent, thin, dielectric, Gaussian phase-randomizing atmospheres, each modeled by a quadratic-phase structure function, with the same coherence length. Being interested only in two OAM states for each photon (those with $\pm l$, where $l$ is fixed at the input), we treated each photon as a qubit, and considered detection of light only in the same two states that were present at the input. Therefore the measurements do not provide a complete characterization of the light. We examined the decay of qubit entanglement by calculating the concurrence using the transmitted density matrix, as a function of the ratio of the optical beam waist to the characteristic turbulence length scale. We assumed that the atmosphere is unmonitored, so any independent information about its fluctuations is lost, leading to loss of entanglement. We found that for a beam waist that is much smaller than the turbulence length, the qubit entanglement is more robust to the turbulent atmosphere. The results also indicated that entangled states with larger OAM values experience less disentanglement through a turbulent atmosphere. The calculation also predicted the sudden disappearance ("death") of entanglement, which was pointed out in the context of two coupled qubits or two atoms in [78, 79].

## 10. Conclusions and Discussion

We have reviewed the Titulaer-Glauber polychromatic, wave-packet field quantization approach to electromagnetism. We showed that the wave-packet modes, which are typically non-orthogonal under the standard scalar product, can be viewed as being orthonormal by defining a new non-local scalar product. This non-local scalar product arises naturally in the context of photon wave mechanics as the Lorentz-invariant scalar product between single photon wave functions. In discussing the non-local scalar product, we used the concept of the dual-mode basis, whose elements are pair-wise orthonormal to the wave-packet modes under the standard overlap integral.

The theory of photon wave mechanics, in which a single excitation of the electromagnetic field is treated as a particle-like entity, has also been reviewed and extended. We have demonstrated how one can derive the photon wave function and its equation of motion by paralleling the Dirac treatment of electrons. Single-photon wave mechanics is shown to be equivalent (in free space) to the standard classical Maxwell theory of electromagnetism in vacuum. When this theory is generalized to multiphoton states, the equivalence between classical coherence theory and photon wave mechanics becomes clear. The $n$ th order coherence tensors evolve in the same manner as the $n$-photon wave functions.

In the general $n$-photon treatment of photon wave mechanics we find that the photon wave functions are identified with the spatial-temporal mode functions of the electromagnetic field. Indeed this connection has been made by others [10, 18, 19, 43, 49]. Yet until now, no one has made the connection with the non-local scalar product within this context, for example, to calculate reduced density matrix elements as we showed. The approach used in the standard treatment of quantum optics, in which the standard scalar product is used, works fine in the quasi-monochromatic limit [4]. However, when broadband photons are involved one must either perform all calculations in the momentum representation, which is not very useful for localized interactions, or, as done here, use the non-local scalar product to calculate the overlap between photon states.

We presented an example of beam-like wave-packet states in which one can expand the photon wave functions. These paraxial-Hermite-Gauss wave functions are approximately orthogonal under the non-local scalar product. To convert between wave-packet modes and their corresponding dual modes, we introduced the concept of a mode converter, which is based upon spectral filtering in the paraxial regime. In addition, we discussed the use of photon wave mechanics to describe the disentanglement of photons entangled in their orbital angular momentum degrees of freedom.

One must take care not to attribute or associate the wave function, density matrix, or state vector in Hilbert space with a quantum object itself, be it a photon, electron, or other fundamental particle. We note that there are three common interpretations of the photon. One is the notion of a fundamental particle. Another is that of an excitation of a quantum field. A third is simply what is registered by a photodetector. In this paper, we do not adhere to any of these as being more correct, as the mathematics of wave functions has little to say on this interpretational matter. A common interpretation of the wave function, density matrix, or state of a quantum object is that it simply gives the maximum available information from which one can calculate the probabilities of experimental outcomes. Ensembles of experimental outcomes can be used to determine the quantum state of identically prepared systems, including photons, as developed in quantum-state tomography [52-54, 80].

Open questions remain in photon wave mechanics. It is not yet clear how to formulate a covariant Lagrangian for the photon wave function, although it seems likely that the approach in [36] will prove useful. A Lagrangian formalism would make it clear how to incorporate interactions of photons with charged particles at a fundamental level. It is not yet clear what experiments might be done in which the non-local scalar product would predict results significantly different from those predicted using the standard overlap integral as an approximate, non-invariant scalar product. It seems likely that an experiment involving fourth-order interference of independently produced ultrashort photons or biphotons would provide such an example.

**Acknowledgements**
We thank Iwo Bialynicki-Birula, Cody Leary, Steven Hsu, Jan Mostowski, John Sipe, Davison Soper, and Ian Walmsley for helpful discussions. BJS was supported by the Royal Society, and both authors were supported by the National Science Foundation, Grant Nos. 0554842, 0456974 and No. 0334590.

**Appendix A. Dual-mode basis in terms of the wave-packet modes**

The dual modes $\mathbf{v}^D_{j,\sigma}(\mathbf{x},t)$, defined in (19) as an integral of the monochromatic plane-wave momentum-space modes $\mathbf{u}_{\mathbf{k},\sigma}(\mathbf{x})$, can be expressed as a spatial integral of the non-monochromatic spatial-temporal wave-packet modes $\mathbf{v}_{j,\sigma}(\mathbf{x},t)$ given in (22). To see this we take the Fourier transform of the wave-packet modes to give the corresponding unitary operators multiplied by the unit polarization vector $U_j^{(\sigma)}(\mathbf{k})\mathbf{e}_{\mathbf{k},\sigma}$, as in (11). This result, (11), is inserted into the definition of the dual modes, (19), to give

$$\begin{aligned}\mathbf{v}^D_{j,\sigma}(\mathbf{x},t) &= i\left(\frac{2\varepsilon_0}{\hbar c}\right)^{1/2}\int \frac{d^3k}{(2\pi)^3}\frac{1}{\sqrt{k}}U_j^{(\sigma)}(\mathbf{k})\mathbf{u}_{\mathbf{k},\sigma}(\mathbf{x})\exp(-i\omega_\mathbf{k} t) \\ &= \frac{2\varepsilon_0}{\hbar c}\int d^3x'\,\mathbf{v}_{j,\sigma}(\mathbf{x}',t)\int\frac{d^3k}{(2\pi)^3}\frac{1}{k}\exp\left[i\mathbf{k}\cdot(\mathbf{x}-\mathbf{x}')\right] \\ &= \int d^3x'\,\mathbf{v}_{j,\sigma}(\mathbf{x}',t)K(\mathbf{x}-\mathbf{x}').\end{aligned} \quad \text{(A1)}$$

Here we have used the Fourier transform pair $1/|\mathbf{k}|$ and $1/|\mathbf{x}|^2$, and introduced the kernel function

$$K(\mathbf{x}) = \frac{2\varepsilon_0}{\hbar c}\int\frac{d^3k}{(2\pi)^3}\frac{1}{|\mathbf{k}|}\exp(i\mathbf{k}\cdot\mathbf{x}) = \frac{\varepsilon_0}{\hbar c\pi^2}\frac{1}{|\mathbf{x}|^2}. \quad \text{(A2)}$$

**Appendix B. Two possible momentum-space wave function normalizations**

In the literature there are two main normalizations used for the momentum-space wave functions $\tilde{\psi}_\sigma(\mathbf{p})$. The most obvious way in which to normalize such wave functions is to unity as we have done in the main text, (40). Here the wave function has the interpretation of being the *probability* amplitude to find a photon with helicity $\sigma$ and momentum between $\mathbf{p}$ and $\mathbf{p}+d\mathbf{p}$ [11]. However, there is another common, but non-standard, normalization used in the literature [8-10]. There the authors normalize the momentum-space wave functions to the average photon energy rather than unity, that is,

$$\sum_\sigma \int \frac{d^3p}{(2\pi\hbar)^3}\tilde{\psi}^{BB}_\sigma(\mathbf{p})^\dagger \tilde{\psi}^{BB}_\sigma(\mathbf{p}) = \langle H \rangle. \quad \text{(B1)}$$

Here $\langle H \rangle$ is the expectation value of the single-photon Hamiltonian, and the superscript *BB* refers to Bialynicki-Birula, the main proponent of this momentum-space wave function normalization. We see that the Bialynicki-Birula momentum-space wave functions have the interpretation of being *energy* amplitudes, instead of probability amplitudes. This implies that $\tilde{\psi}^{BB}_\sigma(\mathbf{p})^\dagger \tilde{\psi}^{BB}_\sigma(\mathbf{p})d^3p/(2\pi\hbar)^3$ is the energy density in momentum space rather than the probability density. The main reason that we can see for this choice of normalization is that the coordinate-space wave functions then become direct Fourier transforms of these energy amplitudes, rather than resorting to a weighted Fourier transform. As pointed out in the text above, the Bialynicki-Birula momentum-space wave functions and the standard momentum-space wave functions are simply related in momentum space through multiplication by the square-root of the monochromatic energy $\sqrt{c|\mathbf{p}|}$, that is,

$$\tilde{\psi}^{BB}_\sigma(\mathbf{p}) = \sqrt{c|\mathbf{p}|}\,\tilde{\psi}_\sigma(\mathbf{p}). \quad \text{(B2)}$$

**Appendix C. Lorentz invariance of scalar-product kernel function**

The kernel function $G(\mathbf{x}-\mathbf{x}')$ defined in (54) is a special case of the more general two-time kernel function

$$J(\mathbf{x}-\mathbf{x}',t-t') = \int \frac{d^3k}{(2\pi)^3} \frac{1}{|\mathbf{k}|} \exp\left[i\mathbf{k}\cdot(\mathbf{x}-\mathbf{x}') - ic|\mathbf{k}|(t-t')\right]. \tag{C1}$$

This kernel can be evaluated in the following way. First write the integral (C1), in spherical coordinates

$$J(\mathbf{x}-\mathbf{x}',t-t') = \frac{1}{(2\pi)^3} \int_0^\infty k^2 dk \int_0^\pi \sin(\theta) d\theta \int_0^{2\pi} d\phi \frac{1}{k} \exp\left[ikr\cos(\theta) - ick\tau\right], \tag{C2}$$

where $r=|\mathbf{x}-\mathbf{x}'|$ and $\tau = t-t'$. The $\phi$ integral gives a factor of $2\pi$, while the $\theta$ integral can be written as

$$\int_0^\pi \sin(\theta) d\theta \exp\left[ikr\cos(\theta) - ick\tau\right] = \int_{-1}^1 dx \exp\left[ikrx - ick\tau\right]$$
$$= \frac{1}{ikr}\left\{\exp\left[ik(r-c\tau)\right] - \exp\left[ik(r+c\tau)\right]\right\}. \tag{C3}$$

The kernel function is thus given by the following integral expression

$$J(\mathbf{x}-\mathbf{x}',t-t') = \frac{1}{4\pi^2 ir} \int_0^\infty dk\left\{\exp\left[ik(r-c\tau)\right] - \exp\left[ik(r+c\tau)\right]\right\}. \tag{C4}$$

This integral can be evaluated using regularization techniques to give

$$J(\mathbf{x}-\mathbf{x}',t-t') = \frac{1}{4\pi^2 ir}\left(\frac{i}{r-c\tau} - \frac{-i}{r+c\tau}\right) = \frac{1}{2\pi^2\left(r^2 - c^2\tau^2\right)}. \tag{C5}$$

This is the generalized kernel function for arbitrary space-time coordinates $(\mathbf{x},t)$ and $(\mathbf{x}',t')$. Notice that it is indeed Lorentz invariant (space-time separations $\Delta(x-x')^2 = |\mathbf{x}-\mathbf{x}'|^2 - c^2(t-t')^2$, are Lorentz invariant). Also note that the case in which $t=t'$, i.e., $\tau = 0$ is the special case discussed above (54).

**Appendix D. Lorentz transformation properties of momentum-space wave functions**

For a normalized, single-photon wave function, as in (40), we require that the norm be Lorentz invariant. By assuming that the form of the norm, (40), is also Lorentz covariant we can determine the Lorentz transformation properties of the momentum space wave functions. Transforming (40) to another inertial frame by a Lorentz transformation $\Lambda$, gives the following norm for the photon wave function in the new frame

$$(\psi\|\psi) = (\psi'\|\psi') = \sum_\sigma \int \frac{d^3 p'}{(2\pi\hbar)^3} \tilde{\psi}_\sigma'(\mathbf{p}')^\dagger \tilde{\psi}_\sigma'(\mathbf{p}') = 1. \tag{D1}$$

Here we use a prime to denote the Lorentz transformed variable. By noting that we may rewrite the integrand in terms of the Lorentz invariant momentum-space volume element $d^3p'/p' = d^3p/p$, we see that

$$\sum_\sigma \int \frac{d^3 p'}{(2\pi\hbar)^3} \frac{p'}{p'} \tilde{\psi}_\sigma'(\mathbf{p}')^\dagger \tilde{\psi}_\sigma'(\mathbf{p}') = \sum_\sigma \int \frac{d^3 p}{(2\pi\hbar)^3} \frac{p'}{p} \tilde{\psi}_\sigma'(\mathbf{p}')^\dagger \tilde{\psi}_\sigma'(\mathbf{p}') = 1. \tag{D2}$$

For this to be Lorentz invariant the modulus squared of the momentum space wave function must transform as

$$\tilde{\psi}_\sigma'(\mathbf{p}')^\dagger \tilde{\psi}_\sigma'(\mathbf{p}') = \frac{p}{p'} \tilde{\psi}_\sigma(\mathbf{p})^\dagger \tilde{\psi}_\sigma(\mathbf{p}) \tag{D3}$$

Thus we may infer that the photon wave function in momentum space transforms as

$$\tilde{\psi}_\sigma'(\mathbf{p}') = \sqrt{\frac{p}{p'}} U(\Lambda, \mathbf{p}, \sigma) \tilde{\psi}_\sigma(\mathbf{p}) \tag{D4}$$

where $U$ is a unitary transformation dependent upon the specific Lorentz transformation $\Lambda$. For details see [10, 58].